\documentclass[final,3p,times]{elsarticle}

\usepackage{graphicx}
\usepackage{amssymb}
\usepackage{amsthm}
\usepackage{color}
\usepackage{float}
\usepackage{verbatim}
\biboptions{sort&compress}

\usepackage{amsmath}
\usepackage{hyperref}
\usepackage{url}
\usepackage{cases}
\usepackage{pdfpages}
\usepackage{cases}
\usepackage{longtable}
\newtheorem{theorem}{Theorem}

\newtheorem{definition}{Definition}

\newtheorem{remark}{Remark}

\DeclareMathOperator{\rg}{rank}

\usepackage{lineno}
\usepackage{color}

\journal{arxiv}

\begin{document}

\begin{frontmatter}


\title{Structural Identifiability and Observability of Compartmental Models of the COVID-19 Pandemic\tnoteref{funding}}

\tnotetext[funding]{This research has received funding from the Spanish Ministry of Science, Innovation and Universities and the European Union FEDER under project grant SYNBIOCONTROL (DPI2017-82896-C2-2-R) and the CSIC intramural project grant MOEBIUS (PIE 202070E062). The funding bodies played no role in the design of the study, the collection and analysis of the data or in the writing of the manuscript.}



\author{Gemma Massonis}\ead{gmassonis@iim.csic.es}
\author{Julio R. Banga}\ead{julio@iim.csic.es}
\author{Alejandro F. Villaverde}\ead{afvillaverde@iim.csic.es}
\address{BioProcess Engineering Group, IIM-CSIC, Vigo 36208, Galicia, Spain}

\begin{abstract}
The recent coronavirus disease (COVID-19) outbreak has dramatically increased the public awareness and appreciation of the utility of dynamic models. At the same time, the dissemination of contradictory model predictions has highlighted their limitations. If some parameters and/or state variables of a model cannot be determined from output measurements, its ability to yield correct insights -- as well as the possibility of controlling the system -- may be compromised. Epidemic dynamics are commonly analysed using compartmental models, and many variations of such models have been used for analysing and predicting the evolution of the COVID-19 pandemic. In this paper we survey the different models proposed in the literature, assembling a list of 36 model structures and assessing their ability to provide reliable information. We address the problem using the control theoretic concepts of structural identifiability and observability. Since some parameters can vary during the course of an epidemic, we consider both the constant and time-varying parameter assumptions.
We analyse the structural identifiability and observability of all of the models, considering all plausible choices of outputs and time-varying parameters, which leads us to analyse 255 different model versions. We classify the models according to their structural identifiability and observability under the different assumptions and discuss the implications of the results. We also illustrate with an example several alternative ways of remedying the lack of observability of a model. Our analyses provide guidelines for choosing the most informative model for each purpose, taking into account the available knowledge and measurements. 
\end{abstract}


\end{frontmatter}


\section{Introduction}\label{sec:intro}

The current coronavirus disease (COVID-19) pandemic, caused by the SARS-CoV-2 virus, continues to wreak unparalleled havoc across the world. Public health authorities can use mathematical models to answer critical questions related with the dynamics of an epidemic (severity and time course of infected people), its impact on the healthcare system, and the design and effectiveness of different interventions \cite{lofgren2014opinion,li2018introduction,currie2020simulation,adam2020special}.
Mathematical modeling of infectious diseases has a long history \cite{Brauer2008book,martcheva2015introduction}. Modeling efforts are particularly important in the context of COVID-19 because its dynamics can be particularly complex and counter-intuitive due to the uncertainty in the transmission mechanisms, possible seasonal variation in both susceptibility and transmission, and their variation within subpopulations \cite{cobey2020modeling}. The media has given extensive coverage to analyses and forecasts using COVID-19 models, with increased attention to cases of conflicting conclusions, giving the impression that epidemiological models are unreliable or flawed. However, a closer looks reveals that these modeling studies were following different approaches, handling uncertainty differently, and ultimately addressing different questions on different time-scales \cite{holmdahl2020wrong}. 

Broadly speaking, data-driven models (using statistical regression or machine learning) can be used for short-term forecasts (one or a few weeks). Mechanistic models based on assumptions about transmission and immunity try to mimic how the virus spreads, and can be used to formalize current knowledge and explore long-term outcomes of the pandemic and the effectiveness of different interventions. However, the accuracy of mechanistic models is constrained by the uncertainties in our knowledge, which creates uncertainties in model parameters and even in the model structure \cite{holmdahl2020wrong}. Further, the uncertainty in the COVID-19 data and the exponential spread of the virus amplify the uncertainty in the predictions.
Predictability studies \cite{scarpino2019predictability} seek the characterization of the fundamental limits to outbreak prediction and their impact on decision-making. Despite the vast literature on mathematical epidemiology in general, and modeling of COVID-19 in particular, comparatively few authors have considered the predictability of infectious disease outbreaks \cite{scarpino2019predictability,castro2020predictability}. Uncertainty quantification \cite{arriola2009sensitivity} is an interconnected concept that is also key for the reliability of a model, and that has received similarly scant attention \cite{faranda2020asymptotic,raimundez2020covid}.

In addition to predictability and uncertainty quantification approaches, identifiability is a related property whose absence can severely limit the usefulness of a mechanistic model \cite{chowell2017fitting}. A model is identifiable if we can determine the values of its parameters from knowledge of its inputs and outputs. Likewise, the related control theoretic property of observability describes if we can infer the model states from knowledge of its inputs and outputs. 
If a model is non-identifiable (or non-observable) different sets of parameters (or states) can produce the same predictions or fit to data. The implications can be enormous: in the context of the COVID-19 outbreak in Wuhan, non-identifiability in model calibrations was identified as the main reason for wide variations in model predictions \cite{RODA2020Why_preddict}. 

Reliable models can be used in combination with optimization and optimal control methods to find the best intervention strategies, such as lock-downs with minimum economic impact \cite{alvarez2020lockdown,acemoglu2020multi}. Further, they can be used to explore the feasibility of model-based real-time control of the pandemic \cite{casella2020can,kohler2020robust}. However, using calibrated models with non-identifiability or non-observability issues can result in bad or even dangerous intervention and control strategies. 

It is common to distinguish between structural and practical identifiability. Structural non-identifiability may be due to the model and measurement (input-output) structure. Practical non-identifiability is due to lack of information in the considered data-sets. Non-identifiability results in incorrect parameter estimates and bad uncertainty quantification \cite{chowell2017fitting, kao-eisenberg-2018practical}, i.e. a misleading calibrated model which should not be used to analyze epidemiological data, test hypothesis, or design interventions. The structural identifiability of several epidemic mechanistic models has been studied e.g. in 
\cite{evans2002structural,evans2005structural,ChapmanEvans2009,eisenberg2013identifiability,brouwer2019integrating,Prague2020Populationmodeling}. Other recent studies have mostly focused on practical identifiability, such as \cite{tuncer2018structural,chowell2017fitting, kao-eisenberg-2018practical,roosa2019assessing,Alahmadi2020}.

In this paper we assess the structural identifiability and observability of a large set of COVID-19 mechanistic models described by deterministic ordinary differential equations, derived by different authors using the compartmental modeling framework \cite{Brauer2008chap}. Compartmental models are widely used in epidemiology because they are tractable and powerful despite their simplicity. 
We collect 36 different compartmental models, of which we consider several variations, making up a total of 255 different model versions.
Our aim is to characterize their ability to provide insights about their unknown parameters -- i.e. their structural identifiability -- and unmeasured states -- i.e. their observability. 
To this end we adopt a differential geometry approach that considers structural identifiability as a particular case of nonlinear observability, allowing to analyse both properties jointly. 
We define the relevant concepts and describe the methods used in Section \ref{sec:methods}. Then we provide an overview of the different types of compartmental models found in the literature in Section \ref{sec:models}. We analyse their structural identifiability and observability and discuss the results in Section \ref{sec:results}, where we also show different ways of remedying lack of observability using an illustrative model. Finally, we conclude our study with some key remarks in Section \ref{sec:conclusions}.

\section{Methods}\label{sec:methods}

\subsection{Notation, models, and properties}

We consider models defined by systems of ordinary differential equations with the following notation:

\begin{numcases}{\mathcal{M}=}\label{din}
	\dot{x}(t)=f\left(x(t),\theta,u(t),w(t)\right)\\\label{out}
	y(t)=h\left(x(t),\theta,u(t),w(t)\right)
\end{numcases}
where $f$ and $h$ are analytical (generally nonlinear) functions of the states $x(t)\in\mathbb{R}^{n_x},$ known inputs $u(t)\in\mathbb{R}^{n_u}$, unknown constant parameters $\theta\in\mathbb{R}^{n_\theta}$, and unknown inputs or time-varying parameters $w(t)\in\mathbb{R}^{n_w}$. The output $y(t)\in\mathbb{R}^{n_y}$ represents the measurable functions of model variables.
The expressions (\ref{din}--\ref{out}) are sufficiently general to represent a wide range of model structures, of which compartmental models are a particular case.

\begin{definition}[Structurally locally identifiable \cite{distefano2015dynamic}]
A parameter $\theta_i$ of model $\mathcal{M}$ is structurally locally identifiable (s.l.i.) if for almost any parameter vector $\theta^*\in\mathbb{R}^{n_\theta}$ there is a neighbourhood ${\mathcal N}(\theta^*)$ in which the following relationship holds:
\begin{equation}\label{eq:sli}
	\hat{\theta} \in {\mathcal N}(\theta^*) \text{ and } y(t,\hat{\theta}) = y(t,\theta^*) \Rightarrow \hat{\theta_i} = \theta_i^*
\end{equation}
Otherwise, $\theta_i$ is \textit{structurally unidentifiable} (s.u.).
If all model parameters are s.l.i. the model is s.l.i. If there is at least one s.u. parameter, the model is s.u..    
\end{definition}

Likewise, a state $x_i(\tau)$ is \textit{observable} if it can be distinguished from any other states in a neighbourhood  
from observations of the model output $y(t)$ and input $u(t)$ in the interval $t_0 \leq \tau \leq t \leq t_f$, for a finite $t_f$. Otherwise, $x_i(\tau)$ is \textit{unobservable}. A model is called observable if all its states are observable. We also say that $\mathcal{M}$ is \textit{invertible} if it is possible to infer its unknown inputs $w(t)$, and we say that $w(t)$ is reconstructible in this case. 

Structural identifiability can be seen as a particular case of observability \cite{tunali1987new,sedoglavic2002probabilistic,villaverde2019observability}, by augmenting the state vector with the unknown parameters $\theta$, which are now considered as state variables with zero dynamics, ${\widetilde x}=(x^T,\theta^T)^T$. The reconstructibility of unknown inputs $w(t)$, which is also known as input observability, can also be cast in a similar way, although in this case their derivatives may be nonzero.
To this end, let us augment the state vector further with $w$ as additional states, as well as their derivatives up to some non-negative integer $l$:
\begin{align}\label{x_aug}
&{\widetilde x}=\begin{pmatrix}
x^T&
\theta^T&
w^T&
\dots&
w^{\left.l\right)^T}\end{pmatrix}^T,
\end{align}
The $l-$augmented dynamics is:
\begin{align*}
&\dot {\widetilde x}(t)=f^l\left(x(t),u(t),w^{\left.l+1\right)}(t)\right)=\begin{pmatrix}
f\left(x(t),u(t)\right)^T&
0_{1\times n_p}&
w(t)^T&
\dots&
w^{\left.l+1\right)}(t)^T
\end{pmatrix}^T,
\end{align*}
leading to the $l-$augmented system:
\begin{align}\label{aug_sys}
\mathcal{M}^l
&\begin{cases}
&\dot {\widetilde x}(t)=f^l(x(t),u(t),w^{\left.l+1\right)}(t))\\
&y(t)=h(x(t),u(t))
\end{cases}
\end{align}

\begin{remark}[Unknown inputs, disturbances, or time-varying parameters]\label{remark_time}
In Section \ref{sec:results}, when reporting the results of the structural identifiability and observability analyses, we will explicitly consider some parameters as time-varying. In the model structure defined in equations (\ref{din}--\ref{out}) the unknown parameter vector $\theta$ is assumed to be constant. To consider an unknown parameter as time-varying we include it in the ``unknown input'' vector $w(t)$. 
Thus, changing the consideration of a parameter from constant to time-varying entails removing it from $\theta$ and including it in $w(t)$.
The elements of $w(t)$ can be interpreted as unmeasured disturbances or inputs of unknown magnitude or, equivalently, as time-varying parameters. Regardless of the interpretation, they are assumed to change smoothly, i.e. they are infinitely differentiable functions of time. For the analysis of some models it is necessary, or at least convenient, to introduce the mild assumption that the derivatives of $w(t)$ vanish for a certain non-negative integer $s$ (possibly $\left.s=+\infty\right)$, i.e. $w^{\left.s\right)}(t)\neq 0$ and $w^{\left.i\right)}(t)=0$ for all $i>s.$ 
This assumption is equivalent to assuming that the disturbances are polynomial functions of time, with maximum degree equal to $s$ \citep{villaverde2019full}.
\end{remark}

\begin{definition}[Full Input-State-Parameter
	Observability, FISPO \citep{villaverde2019full}]
	Let us consider a model $\mathcal{M}$ 
	given by (\ref{din}--\ref{out}). 
We augment its state vector as $z(t)=\begin{pmatrix}
	x(t)^T&\theta^T&w(t)^T\end{pmatrix}^T$ (\ref{x_aug}), which leads to its augmented form \eqref{aug_sys}. 
We say that $\mathcal{M}$ has the FISPO property if, for every $t_0\in I$, every model unknown 
$z_i(t_0)$ can be inferred from $y(t)$ and $u(t)$ in a finite time interval $\left[t_0,t_f\right]\subset I.$ Thus, $\mathcal{M}$ is FISPO if, for every $z(t_0)$ and for almost any vector $z^\ast(t_0),$ there is a neighbourhood $\mathcal{N}\left(z^\ast\left(t_0\right)\right)$ such that, for all $\hat{z}(t_0)\in\mathcal{N}\left(z^\ast\left(t_0\right)\right),$ the following property is fulfilled:
	\begin{align*}
	&y\left(t,\hat{z}(t_0)\right)=y\left(t,z^\ast\left(t_0\right)\right)\Rightarrow \hat{z}_i\left(t_0\right)=z_i^\ast\left(t_0\right),\quad 1\leq i \leq n_x+n_\theta+n_w.
	\end{align*}
\end{definition}

\subsection{Structural identifiability and observability analysis}

In this paper we analyse input, state, and parameter observability --  that is, the FISPO property defined above -- using a differential geometry framework. Such analyses are \textit{structural} and \textit{local}. By structural we refer to properties that are entirely determined by the model equations; thus we do not consider possible deficiencies due to insufficient or noise-corrupted data. By local we refer to the ability to distinguish between neighbouring states (similarly, parameters or unmeasured inputs), even though they may not be distinguishable from other distant states. This is usually sufficient, since in most (although not all, see e.g. \cite{thomaseth2018local}) applications local observability entails global observability. This specific type of observability has sometimes been called local weak observability \citep{hermann1977nonlinear}. 

This approach assesses structural identifiability and observability by calculating the rank of a matrix that is constructed with Lie derivatives. The corresponding definitions are as follows (in the remainder of this section we omit the dependency on time to simplify the notation):

\begin{definition}[Extended Lie derivative \citep{anguelova2012efficient}]\label{lie_ext}
	Consider the system $\mathcal{M}$ (\ref{din}--\ref{out}) with augmented state vector (\ref{x_aug}) and augmented dynamics \eqref{aug_sys}. Assuming that the inputs $u$ are analytical functions, the extended Lie derivative of the output along $\tilde{f}=\tilde f\left(\cdot,u\right)$ is:
	\begin{align*}
	&L_{\tilde{f}}^h\left(\tilde{x},u\right)=\frac{\partial h}{\partial \tilde{x}}\left(\tilde{x},u\right) \tilde{f}\left(\tilde{x},u\right)+\frac{\partial h}{\partial u}\left(\tilde{x},u\right) \dot{u}.
	\end{align*}
	The zero-order derivative is $L^{0}_{\tilde{f}}h=h,$ and the $i-$order extended Lie derivatives can be recursively calculated as:
	\begin{align*}
	&L^{i}_{\tilde{f}}h\left(\tilde{x},u\right)=\frac{\partial L^{i-1}_{\tilde{f}}h}{\partial \tilde{x}}\left(\tilde{x},u\right) \tilde{f}\left(\tilde{x},u\right)+\sum_{j=0}^{i-1}\frac{\partial L^{i-1}_{\tilde{f}}h}{\partial u^{\left.j\right)}}\left(\tilde{x},u\right) u^{\left.j+1\right)},\quad i\geq 1.
	\end{align*}
\end{definition}

\begin{definition}[Observability-identifiability matrix \cite{villaverde2019full}]
The observability-identifiability matrix of the system $\mathcal{M}$ (\ref{din}--\ref{out}) with augmented state vector (\ref{x_aug}), augmented dynamics \eqref{aug_sys}, and analytical inputs $u$ is the following $mn_{\tilde{x}}\times n_{\tilde{x}}$ matrix,
\begin{align}\label{matr_obs}
&\mathcal{O}_I\left(\tilde{x},u\right)=\frac{\partial}{\partial \tilde{x}}
\begin{pmatrix}
L^0_{\tilde{f}}h\left(\tilde{x},u\right)^T&L_{\tilde{f}}h\left(\tilde{x},u\right)^T&L^2_{\tilde{f}}h\left(\tilde{x},u\right)^T&\dots& L^{n_{\tilde{x}}-1}_{\tilde{f}}h\left(\tilde{x},u\right)^T
\end{pmatrix}^T,
\end{align}
\end{definition}

The FISPO property of $\mathcal{M}$ can be analysed by calculating the rank of the observability-identifiability matrix:

\begin{theorem}[Observability-identifiability condition, OIC \citep{anguelova2012efficient}]\label{OIC}
	If the identifiability-observability matrix of a model $\mathcal{M}$ satisfies $\rg\left(\mathcal{O}_I\left(\tilde{x}_0,u\right)\right)=n_{\tilde{x}} = n_x+n_\theta+n_w,$ with $\tilde{x}_0$ being a (possibly generic) point in the augmented state space, then the system is structurally locally observable and structurally locally identifiable.
\end{theorem}

\subsubsection{Analysis tools}

In this paper we generally check the OIC criterion of (\ref{OIC}) using STRIKE-GOLDD, an open source MATLAB toolbox \cite{villaverde2016structural}. Alternatively, for some models we use the Maple code ObservabilityTest, which implements a procedure that avoids the symbolic calculation of the Lie derivatives and is hence computationally efficient \cite{sedoglavic2002probabilistic}. 
A number of other software tools are available, including GenSSI2 \cite{ligon2018genssi} in MATLAB, IdentifiabilityAnalysis in Mathematica \cite{anguelova2012efficient}, DAISY in REDUCE \cite{saccomani2019new}, SIAN in Maple \cite{hong2018global}, and the web app COMBOS \cite{meshkat2014finding}.
It should be taken into account that in the present work we are interested in assessing structural identifiability and observability both with constant and continuous time-varying model parameters (or equivalently, with unknown inputs), as explained in Remark \ref{remark_time}. Ideally, the method of choice should provide a convenient way of analysing models with this type of parameters (inputs). It is always possible to perform this type of analysis by assuming that the time dependency of the parameters is of a particular form, e.g. a polynomial function of a certain maximum degree.

\section{Models}\label{sec:models}
In this article we review compartmental models, which are one of the most widely used families of models in epidemiology.
They divide the population into homogeneous compartments, each of which corresponds to a state variable that quantifies the number of individuals that are at a certain disease stage. The dynamics of these compartments are governed by ordinary differential equations, usually with unknown parameters that describe the rates at which individuals move among different stages of disease.

The basic compartmental model used for describing a transmission disease is the SIR model, in which the population is divided into three classes:
\begin{itemize}
    \item Susceptible: individuals who have no immunity and may become infected if exposed.
    \item Infected and infectious: an exposed individual becomes infected after contracting the disease. Since an infected individual has the ability to transmit the disease, he/she is also infectious. 
    \item Recovered: individuals who are immune to the disease and do not affect its transmission.
\end{itemize}
Another class of models, called SEIR, include an additional compartment to account for the existence of a latent period after the transmission:
\begin{itemize}
    \item Exposed: individuals vulnerable to contracting the disease when they come into contact with it.
\end{itemize}

These idealized models differ from the reality. Contact tracing, screening, or changes in habits are some differences that are not considered in basic SIR or SEIR models, but are important for evaluating the effects of an intervention. Furthermore, it is not only important to enrich the information about the behaviour of the population; the characteristics of the disease must also be taken into account. These additional details can be incorporated to the model as new parameters, functions, or extra compartments. Compartments such as asymptomatic, quarantined, isolated, and hospitalized have been widely used in COVID-19 models.
From 29 articles, most of which are very recent \cite{zheng2020total,Maier2020effective_containment,RODA2020Why_preddict,Giordano2020SIDARTHE, castro2020predictability,gaeta2020simple,Lourenco2020FUndamental_principles,Raimundez2020Outbreak_wuhan,franco2020feedback,Fosu2020Construction,Ryan2013Github,Lopez2020ModifiedSEIR,Hubbs2020SocialDistancing,pribylova2020seiar,Wen2020OptimalStrategies,Liangrong2020Epidemic_Analysis,sameni2020mathematical,EIKENBERRY2020Mask,Shi2020SEIR_Transmission,CHATTERJEE2020Healthcare,Jia2020Modeling_control,Rahman2020Modelling_transmission,Dohare2020Mathematical_Model,Gevertz2020NovelCOVID}, we have collected 36 models. Depending on whether they have an exposed compartment or not, they can be broadly classified as belonging to the SIR or SEIR families. However, most of these models include additional compartments. 

\subsection{SIR models}
Susceptible individuals become infected with an incidence of:
$$-\beta S I $$
where $\beta=p c$ is the transmission rate, $c$ is the contact rate and $p$ the probability that a contact with a susceptible individual results in a transmission \cite{martcheva2015introduction}.
Individuals who recover leave the infectious class at rate $\gamma$, where $1 / \gamma$ is the average infectious period. 
The set of differential equations describing the basic SIR model is given by:
\begin{equation}
    \begin{split}
        \dot{S}(t)&=-\beta S(t) I(t)\\
        \dot{I}(t)&=\beta S(t) I(t)-\gamma R(t)\\
        \dot{R}(t)&=\gamma R(t)
    \end{split}
\end{equation}

\begin{figure}[H]
\centering\includegraphics[width=1\linewidth]{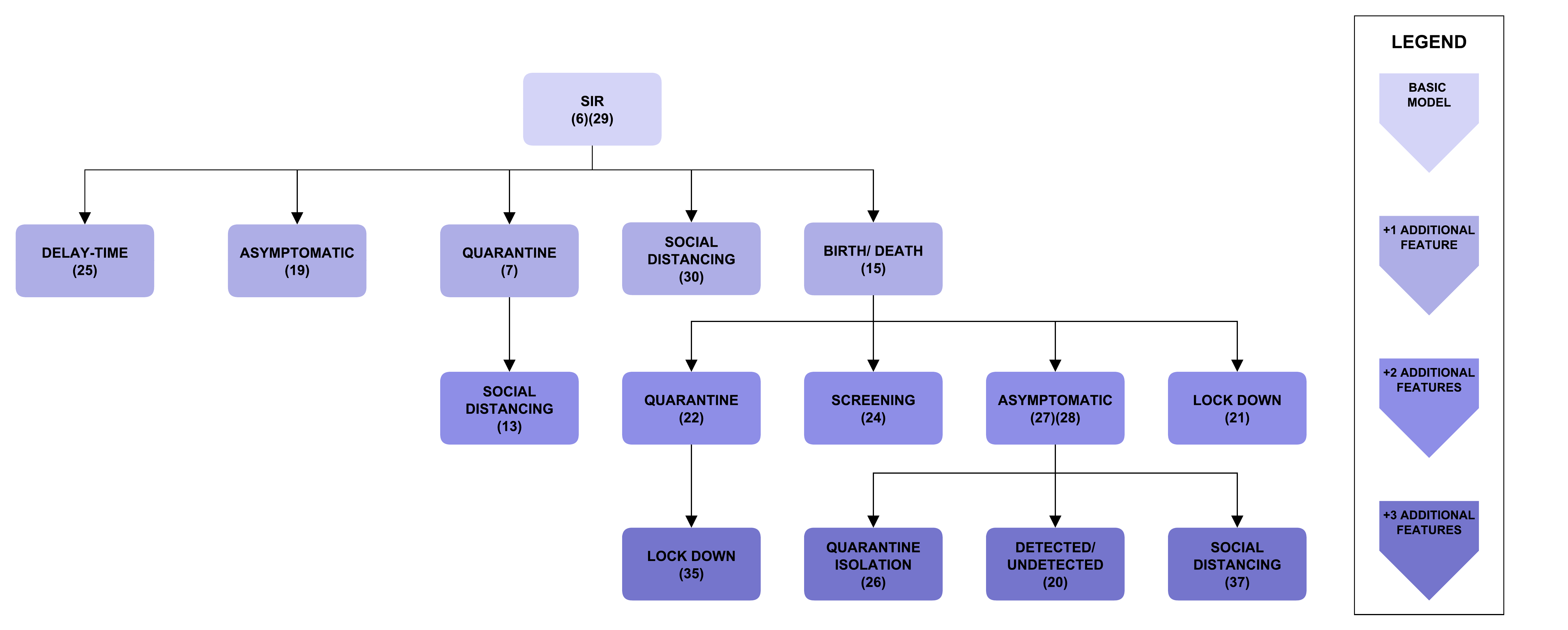}
\caption{Classification of SIR models. Each block represents a model structure. The basic, three-compartment SIR model structure is on top of the tree. Every additional block is labeled with the additional feature that it contains with respect to its parent block. The darkness of the shade indicates the number of additional features with respect to the basic SIR model.}
\label{fig_sir_class}
\end{figure}

As mentioned above, compartmental models can be extended to consider further details. We have found models that incorporate the following features: asymptomatic individuals, births and deaths, delay-time, lock-down, quarantine, isolation, social distancing, and screening.
Figure \ref{fig_sir_class} shows a classification of the SIR models reviewed in this article, and Table \ref{List_SIR_models} lists them along with their equations. Multiple output choices have been considered in the study of the structural identifiability and observability of some models. In such cases the observations are listed in the \textit{Output} column.


{\footnotesize
\begin{longtable}{l l l l l l l l}
\hline
\textbf{ID} & \textbf{Ref.} & \textbf{States} & \textbf{Parameters} & \textbf{Output} & \textbf{ICS} & \textbf{Input} & \textbf{Equations}\\
\hline
6 & \cite{zheng2020total} & S, I, R & $\beta$, $\gamma$ & \parbox{0.3cm}{\begin{align*}
& (1) I \\
& (2) KI 
\end{align*}} & 
\parbox{1cm}{\begin{align*}
& 100000\\
& 100 \\
& 0
\end{align*}} & &
\parbox{0.3cm}{\begin{align*}
        \dot{S}(t)&= - \beta SI/N\\
        \dot{I}(t)&=\beta SI/N- \gamma I\\
        \dot{R}(t)&=\gamma I
\end{align*}}
\\
 \hline
 7 & \cite{zheng2020total} & S, I, R, Q &  $\beta$, $\gamma$, $\delta$ & \parbox{1cm}{\begin{align*}
& (1) Q\\
& (2) I, R, Q  
\end{align*}} &
\parbox{1cm}{\begin{align*}
 &1000\\
& 10\\
& 0\\
& 0
\end{align*}} & &
\parbox{1cm}{\begin{align*}
        \dot{S}(t)&= - \beta SI/N \\
        \dot{I}(t)&=\beta SI/N- \gamma I-\delta I\\
        \dot{R}(t)&=\gamma I\\
        \dot{Q}(t)&=\delta I
\end{align*}}\\
\hline
 13 & \cite{Maier2020effective_containment} & S, I, R, X &  
 \vtop{
 \hbox{\strut $\beta$, $\alpha$, k,}
 \hbox{\strut k$_0$, C($t_0$)}
 }
   & NX & \parbox{1cm}{\begin{align*} 
   & 1-I(t_0)-X(t_0)\\ 
 & I_0 C(t_0)/(X_0 N)\\
 & C(t_0)/N \\
 & R_0
 \end{align*}}   & &
\parbox{1cm}{\begin{align*} 
\dot{S}(t)&= - \alpha SI -k_0 S\\
        \dot{I}(t)&=\alpha SI- (\beta +k + k_0 )I\\
        \dot{X}(t)&=(k+k_0) I\\
        \dot{R}(t)&=\beta I+k_0 S
  \end{align*}}\\
\hline
 15 & \cite{RODA2020Why_preddict} & S, I, R &  $\beta$, $\rho$, $\mu$, $\tau$, d, I$_0$ & $\rho I +\tau$ &      \parbox{1cm}{\begin{align*}
 & S_0\\
     & I_0 \\
     & R_0
     \end{align*}} & &
\parbox{1cm}{\begin{align*}
        \dot{S}(t)&= - \beta SI\\
        \dot{I}(t)&=\beta SI- (\rho +\mu) I\\
        \dot{R}(t)&=\rho I -d R
    \end{align*}}\\
\hline
 20 & \cite{Giordano2020SIDARTHE} & 
 \vtop{
 \hbox{\strut S, I, D,}
 \hbox{\strut A, R, T,}
 \hbox{\strut H, E}
 }
    & 
 \vtop{\hbox{ \strut $\alpha$, $\beta$, $\rho$, $\mu$, $\tau$}
 \hbox{\strut $\epsilon$, $\eta$, $\xi$, $\lambda$, $\sigma$, $\kappa$,}
 \hbox{\strut $\theta$, $\nu$, $\gamma$, $\delta$, $\zeta$}
 
 }
     & D, R, T & 
\parbox{1cm}{\begin{align*}
     & S_0\\
     & 200/60e8\\
     & 20/60e8\\
     & 1/60e8 \\
     & 2/60e8\\
     & 0\\
     & 0\\
     & 0
     \end{align*}}
      & &
\parbox{1cm}{\begin{align*}
        \dot{S}(t)&= - S(\alpha I+\beta D +\gamma A+\delta R)\\
        \dot{I}(t)&=S(\alpha I+\beta D +\gamma A+\delta R)-
       (\epsilon+\zeta+\lambda)I\\
       \dot{D}(t)&=\epsilon I-(\eta +\rho)D\\
        \dot{A}(t)&=\zeta I -(\theta+\mu +\kappa)A\\
        \dot{R}(t)&=\eta D+\theta A-(\nu +\xi)R\\
        \dot{T}(t)&=\mu A+\nu R-(\sigma +\tau )T\\
        \dot{H}(t)&=\lambda I+\rho D +\kappa A+
        \xi R+\sigma T\\
       \dot{E}(t)&=\tau T
   \end{align*}}\\
\hline
 21 & \cite{castro2020predictability} & 
 \vtop{
 \hbox{\strut S, I, D,}
 \hbox{\strut C, R}
 }
    & 
 \vtop{\hbox{ \strut p, q, r}
 \hbox{\strut $\beta$, $\mu$ }
 
 }
     & \parbox{1cm}{\begin{align*}
& (1) I \\
& (2) C  \\
& (3) C, D
\end{align*}} & 
     
      & &
\parbox{1cm}{\begin{align*}
        \dot{S}(t)&= - \beta SI/N-pS+qC \\
        \dot{I}(t)&=\beta SI/N-(r+\mu)I\\
        \dot{R}(t)&=rI \\
        \dot{C}(t)&=pS-qC \\
        \dot{D}(t)&=\mu I
   \end{align*}}\\
\hline
 19 & \cite{gaeta2020simple} & 
 \vtop{
 \hbox{\strut S, I, J,}
 \hbox{\strut R, U}
 }
    & 
 \vtop{\hbox{ \strut $\beta$, $\alpha$, $\eta$, $\xi$}

 }
     & \parbox{1cm}{\begin{align*}
& (1) R\\
& (2) I, R
\end{align*}} & 
\parbox{1cm}{\begin{align*}
     &S_0 \\
     & I_0\\
    & (1-\xi)I_0/\xi \\
     & R_0 \\
     & U_0
     \end{align*}}
     
      & &
\parbox{1cm}{\begin{align*}
        \dot{S}(t)&= - \alpha (I+J)S\\
        \dot{I}(t)&=\alpha \xi S(I+J)-\beta I\\
        \dot{J}(t)&=\alpha (1-\xi)S(I+J)-\eta J\\
        \dot{R}(t)&=\beta I\\
        \dot{U}(t)&=\eta J
   \end{align*}}\\
\hline
 24 & \cite{KIM199Mathematical_model} & 
 \vtop{
 \hbox{\strut S, I, R}
 }
    & 
 \vtop{\hbox{ \strut $\mu$, $\gamma$, $\phi$}
\hbox{ \strut c, K}
 
 }
     & KI & 
      &$\sigma(t)$ &
\parbox{1cm}{\begin{align*}
        \dot{S}(t)&= \Lambda -\mu S-c \phi \dfrac{SI}{S+I}\\
        \dot{I}(t)&=-\mu I +c \phi \dfrac{SI}{S+I}-\gamma I-
        I \dfrac{\sigma(S+I)}{S+I}\\
        \dot{R}(t)&=-\mu R+\gamma I+ I \dfrac{\sigma(S+I)}{S+I}\\
   \end{align*}}\\
\hline
 25 & \cite{Lourenco2020FUndamental_principles} & 
 \vtop{
 \hbox{\strut y, z, A}
 }
    & 
 \vtop{\hbox{ \strut $\beta$, $\sigma$, }
\hbox{ \strut $\rho$, $\theta$}
 }     & A & 
     \parbox{1cm}{\begin{align*}
      & y_0\\
& z_0 \\
&A_0
\end{align*}}

      &z$_d$ &
\parbox{1cm}{\begin{align*}
        \dot{y}(t)&= \beta y(1-z)-\sigma y\\
        \dot{z}(t)&=\beta y(1-z)\\
        \dot{A}(t)&=N \rho \theta z_d
   \end{align*}}\\
\hline
 26 & \cite{Gallina2012epidemic_spreads} & 
 \vtop{
 \hbox{\strut S, I, R,}
  \hbox{\strut A, Q, J}
 }
    & 
 \vtop{\hbox{ \strut $\mu_1$, $\mu_2$, d$_1$, d$_2$,}
\hbox{ \strut d$_3$, d$_4$, d$_5$, d$_6$,}
 \hbox{ \strut k$_1$, k$_2$, $\lambda$, $\gamma_1$,}
 \hbox{ \strut $\gamma_2$, $\epsilon_a$, $\epsilon_q$, $\epsilon_j$}

 }
     & Q, J &   & &
\parbox{1cm}{\begin{align*}
        \dot{S}(t)&= bN-S(I\lambda +\lambda Q\epsilon_a \epsilon_q+
        \lambda \epsilon_a A+\lambda \epsilon_j J+d_1)\\
        \dot{I}(t)&=k_1 A-(\gamma_1+\mu_2+d_2)I\\
        \dot{R}(t)&=\gamma_1 I+\gamma_2 J-d_3 R\\
        \dot{A}(t)&=S(I\lambda +\lambda Q\epsilon_a \epsilon_q+\lambda \epsilon_a A+
        \lambda \epsilon_j J)-(k_1+\mu_1+d_4)A\\
        \dot{Q}(t)&=\mu_1 A-(k_2+d_5)Q\\
        \dot{J}(t)&=k_2Q+\mu_2 I -(\gamma_2+d_6)J
   \end{align*}}\\
\hline
 22 & \cite{HETHCOTE2002Effects_quarantine} & 
 \vtop{
 \hbox{\strut S, I, Q,}
  \hbox{\strut R}
 }
    & 
 \vtop{\hbox{ \strut  d, $\epsilon$, $\beta$}
\hbox{ \strut $\gamma$, $\delta$, $\alpha_1$, $\alpha_2$}
 }
     & Q &   & &
\parbox{1cm}{\begin{align*}
        \dot{S}(t)&= A- \beta SI-dS\\
        \dot{I}(t)&=\beta S I-I(\gamma +d+\delta+\alpha_1)\\
        \dot{Q}(t)&=\delta I-(\epsilon+d+\alpha_2)Q\\
        \dot{R}(t)&=\gamma I+\epsilon Q- R
\end{align*}}\\
\hline
 27 & \cite{Raimundez2020Outbreak_wuhan} & 
 \vtop{
 \hbox{\strut S, A, I,}
  \hbox{\strut R, D}
 }
    & 
 \vtop{\hbox{ \strut $\gamma$, $\delta$, k, }
\hbox{ \strut $\zeta_0$, $\beta_0$}
 }
     & \parbox{1cm}{\begin{align*}
&(1) I, R\\
&(2) R, D \\
&(3) I, R, D
\end{align*}} & 
     \parbox{1cm}{\begin{align*}
     &S_0 \\
&A_0\\
 &I_0\\
 &R_0\\
 &D_0
\end{align*}}
     
     &     g(t)&
\parbox{1cm}{\begin{align*}
        \dot{S}(t)&= -\beta_0 gSI/N-\zeta_0 gSA/N\\
        \dot{A}(t)&=\beta_0 gSI/N+\zeta_0 gSA/N-kA\\
        \dot{I}(t)&=kA-(\gamma+\delta)I\\
        \dot{R}(t)&=\gamma I\\
        \dot{D}(t)&=\delta I
\end{align*}}\\
\hline
 28 & \cite{Raimundez2020Outbreak_wuhan} & 
 \vtop{
 \hbox{\strut S, A, I,}
  \hbox{\strut R, RR, D}
 }
    & 
 \vtop{\hbox{ \strut $\gamma$, $\delta$, k, }
\hbox{ \strut $\nu$, $\zeta_0$, $\beta_0$}
 }
     & \parbox{1cm}{\begin{align*}
&(1) I, R\\
&(2) R, D\\
&(3) I, R, D
\end{align*}} & 
     \parbox{1cm}{\begin{align*}
     &S_0\\
& A_0\\
& I_0\\
& R_0\\
& RR_0\\
& D_0\\
\end{align*}}
     &     g(t)&
\parbox{1cm}{\begin{align*}
        \dot{S}(t)&= -\beta_0 gSI/N-\zeta_0 gSA/N\\
       \dot{A}(t)&=\beta_0 gSI/N+\zeta_0 gSA/N-kA\\
        \dot{I}(t)&=kA-(\gamma+\delta)I\\
        \dot{R}(t)&=\gamma I\\
        \dot{RR}(t)&=\nu A \\
        \dot{D}(t)&=\delta I
\end{align*}}\\
\hline
 29 & \cite{franco2020feedback} & 
 \vtop{
 \hbox{\strut s, i}
 }
    & 
 \vtop{\hbox{ \strut R$_0$ }
 }
     & i & 
     \parbox{1cm}{\begin{align*}
     & N- 2/e6 \\
    & 2/e6
\end{align*}}
     
     &     &
        \parbox{1cm}{\begin{align*}
        \dot{s}(t)&= -R_0 si\\
        \dot{i}(t)&=(R_0 s-1)i
\end{align*}}\\
\hline
 30 & \cite{franco2020feedback} & 
 \vtop{
 \hbox{\strut s, i}
 }
    & 
 \vtop{\hbox{ \strut R$_0$, $\kappa$ }
 }
     & i & 
\parbox{1cm}{\begin{align*}
        & N- 2/e6 \\
        &2/e6 
\end{align*}}
     
     &     &
\parbox{1cm}{\begin{align*}
\dot{s}(t)&= -\dfrac{R_0 si}{1+\kappa i}\\
\dot{i}(t)&=\left(\dfrac{R_0 s}{1+\kappa i}-1 \right)i
\end{align*}}\\
\hline
 35 &\cite{Fosu2020Construction} & 
 \vtop{
 \hbox{\strut S, L, I,}
 \hbox{\strut Q, R}
 }
    & 
 \vtop{\hbox{ \strut  $\gamma$, $\beta$, $\eta$, }
 \hbox{ \strut  $\delta$, $\theta_1$, $\alpha_1$, }
 \hbox{ \strut $\alpha_2$ }
 }
     & Q, L &      &     &
\parbox{1cm}{\begin{align*}
        \dot{S}(t)&= \mu N-\beta S I-(\gamma+\eta)S+\delta L\\
        \dot{L}(t)&=\eta S-(\gamma+\delta)L\\
        \dot{I}(t)&=\beta S I-(\gamma +\theta_1+\alpha_1)I\\
       \dot{Q}(t)&=\theta_1 I-(\gamma+\alpha_2)Q\\
        \dot{R}(t)&=\alpha_1 I+\alpha_2 Q-\gamma R
\end{align*}}\\
\hline

37 & \cite{Gevertz2020NovelCOVID} &  \vtop{
 \hbox{\strut Sd, Sn, Ad,}
 \hbox{\strut An, I, R}
 }
 & 
 \vtop{\hbox{ \strut $\epsilon_s$, $\epsilon_i$, f, }
 \hbox{ \strut h$_1$, h$_2$, $\beta_i$, $\delta$, }
 \hbox{ \strut $\beta_a$, $\gamma_{ai}$, $\gamma_{ir}$ }
 }
     & Sd, I &   \parbox{1cm}{\begin{align*}  
     &0\\
     & 1-10^{-5}\\
     &0\\
     &0\\
     &10^{-5}\\
     &0
     \end{align*}}&     &
\parbox{1cm}{\begin{align*}
        \dot{Sd}(t)&= -\epsilon_s \beta_a (A_n+\epsilon_a A_d)S_d -
        h_1 S_d+h_2 S_n-\epsilon_s \beta_i S_d I\\
        \dot{Sn}(t)&=-\beta_i S_n I-\beta_a(A_n+\epsilon_a A_d)S_n+
        h_1 S_d-h_2 S_n\\
        \dot{Ad}(t)&=\epsilon_s \beta_i S_d I+\epsilon_s \beta_a (A_n+\epsilon_a A_d)S_d+
        h_2 A_n -\gamma_{ai} A_d-h_1 A_d\\
       \dot{An}(t)&=\beta_i S_n I+\beta_a (A_n+\epsilon_a A_d)S_n+
       h_1 A_d -\gamma_{ai} A_n -h_2 A_n\\
        \dot{I}(t)&=f \gamma_{ai}(A_d+A_n)-\delta I-\gamma_{ir} I \\
        \dot{R}(t)&=(1-f) \gamma_{ai}(A_d+A_n)+\gamma_{ir}I
\end{align*}}\\
\hline
\caption{List of SIR models and their main features}
\label{List_SIR_models}
\end{longtable}
}
%
%
%

\subsection{SEIR models}
Individuals in the SEIR model are divided in four compartments: Susceptible (S), Exposed (E), Infected (I) and Recovered (R). Compared to the SIR models, the additional compartment $E$ allows for a more accurate description of diseases in which the incubation period and the latent period do not coincide, i.e. the period between which an infected becomes infectious. This is why SEIR models are in principle best suited to epidemics with a long incubation period such as COVID-19 \cite{franco2020feedback}.  

Susceptible individuals move to the exposed class at a rate $\beta I(t)$, where $\beta$ is the transmission rate parameter. Exposed individuals become infected at rate $\kappa$, where $1/\kappa$ is the average latent period. Infected individuals recover at rate $\gamma$, where $1/\gamma$ is the average infectious period.

Thus, the set of differential equations describing the basic SEIR model is:
\begin{equation}
    \begin{split}
        \dot{S}(t)&=-\beta S(t) I(t)\\
        \dot{E}(t)&=\beta S(t) I(t)-\kappa E\\
        \dot{I}(t)&=k E-\gamma R(t)\\
        \dot{R}(t)&=\gamma R(t)
    \end{split}
\end{equation}

\begin{figure}[H]
\centering\includegraphics[width=1.1\linewidth]{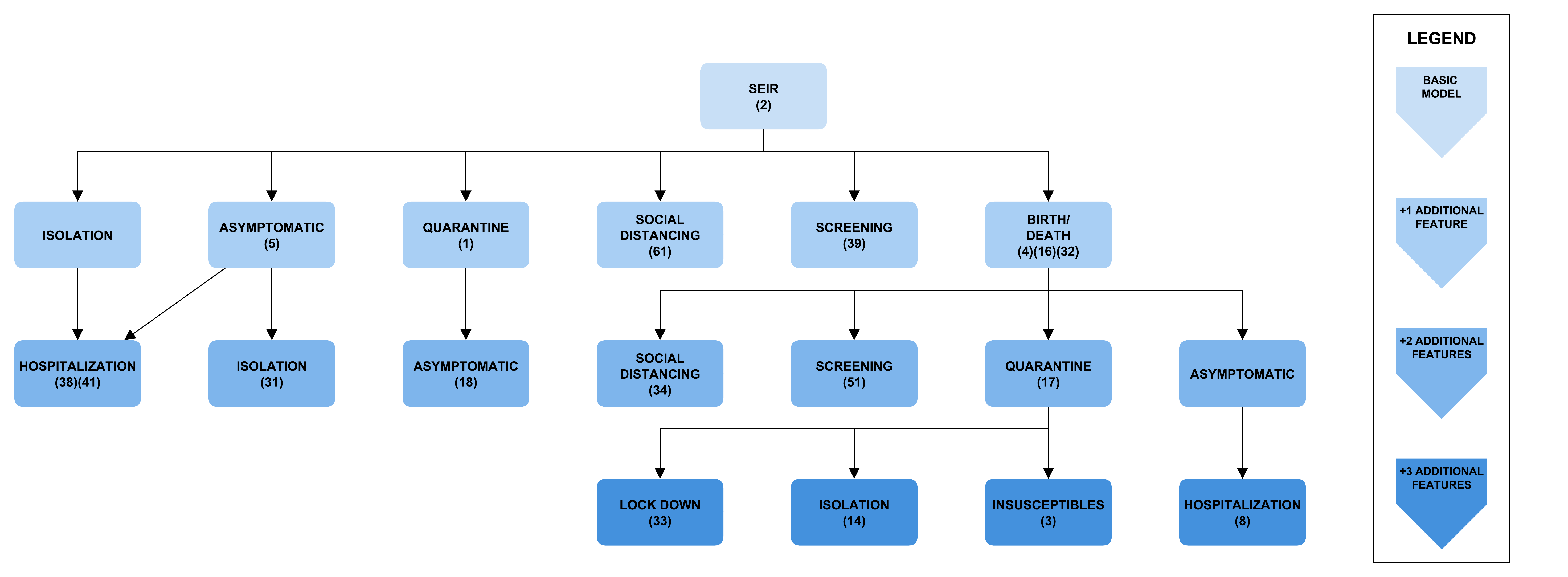}
\caption{Classification of SEIR models. Each block represents a model structure. The basic, four-compartment SEIR model structure is on top of the tree. Every additional block is labeled with the additional feature that it contains with respect to its parent block. The darkness of the shade indicates the number of additional features with respect to the basic SEIR model.}
\label{fig_seir_class}
\end{figure}

Existing extensions of SEIR models may incorporate some of the following features: asymptomatic individuals, births and deaths, hospitalization, quarantine, isolation, social distancing, screening and lock-down. Figure \ref{fig_seir_class} shows a classification of the models found in the literature; Table \ref{List_SEIR_models} lists them along with their equations. 

{\footnotesize
\begin{longtable}{l l l l l l l l}
\hline
\textbf{ID} & \textbf{Ref.} & \textbf{States} & \textbf{Parameters} & \textbf{Output} & \textbf{ICS} & \textbf{Input} & \textbf{Equations}\\
\hline
2 & \cite{roosa2019assessing} & S, E, I, R & 
 $\beta$, $\gamma$, k
 & \parbox{1cm}{\begin{align*}
& (1)C \\
& (2)KC 
\end{align*}} & 
 & &
\parbox{1cm}{\begin{align*}
        \dot{S}(t)&= - \beta SI/N\\
        \dot{E}(t)&=\beta SI/N-kE\\
        \dot{I}(t)&=kE- \gamma I\\
        \dot{R}(t)&=\gamma I \\
        \dot{C}(t) & =kE
\end{align*}}
 \\
 \hline
 34 & \cite{Chitnis2017SEIR} & S, E, I, R & 
 \vtop{
 \hbox{\strut  $\beta$, $\gamma$, $\mu$, }
 \hbox{\strut $\epsilon$, r, K}
 }
 & KI & 
 & &
\parbox{1cm}{\begin{align*}
        \dot{S}(t)&=\Lambda - r\beta SI/N-\mu S\\
        \dot{E}(t)&=\beta SI/N-\epsilon E-\mu E\\
        \dot{I}(t)&=\epsilon E- \gamma I-\mu I\\
        \dot{R}(t)&=\gamma I-\mu R 
\end{align*}}
 \\
 \hline
  16 & \cite{RODA2020Why_preddict} & S, E, I, R & 
 \vtop{
 \hbox{\strut $\beta$, $\rho$, $\mu$, d, }
 \hbox{\strut $\epsilon$, $\tau$, $E_0$, $I_0$}
 }
 & 
 \parbox{1cm}{\begin{align*}
&(1) \rho I+\tau\\
&(2) \mu I 
\end{align*}}& 
 \parbox{1cm}{\begin{align*}
        & S_0\\
        &E_0\\
        &I_0\\
        & R_0 
\end{align*}}
 & &
\parbox{1cm}{\begin{align*}
        \dot{S}(t)&=-\beta SI\\
        \dot{E}(t)&=\beta SI-\epsilon E\\
        \dot{I}(t)&=\epsilon E- (\rho+\mu) I\\
        \dot{R}(t)&=\rho I-d R 
\end{align*}}
 \\
 \hline
   51 & \cite{Ryan2013Github} & \vtop{
 \hbox{\strut S, E, I, De, }
 \hbox{\strut Di, R, F}
 } & 
 \vtop{
 \hbox{\strut  q, $\mu_i$, $\mu_d$, }
 \hbox{\strut $\sigma$, $\sigma_d$, $\theta_e$, $\theta_i$, }
 \hbox{\strut $\gamma$, $\gamma_d$, $\beta$, $\beta_d$, }
 \hbox{\strut $\phi_i$, $\phi_e$, $\mu_0$ }
 }
 &  \parbox{1cm}{\begin{align*}
&(1) De, Di, F\\
&(2) Di, F 
\end{align*}} &  & &
\parbox{1cm}{\begin{align*}
        \dot{S}(t)&=-\beta SI/N-q \beta_d S Di/N+
        \nu N -\mu_0 S\\
        \dot{E}(t)&=\beta SI/N+q \beta_d S Di/N-
        \sigma E-\theta_e \phi_e E-\mu_0 E\\
        \dot{I}(t)&=\sigma E- \gamma I-\mu_i I-
        \theta_i \phi_i I-\mu_0 I\\
        \dot{De}(t)&=\theta_e \phi_e E-\sigma_d De-\mu_0 De\\
        \dot{Di}(t)&=\theta_i \phi_i I+\sigma_d De-\gamma_d Di -
        \mu_d Di-\mu_0 Di\\
        \dot{R}(t)&=\gamma I+\gamma_d Di-\mu_0 R\\
        \dot{F}(t)&=\mu_i I+\mu_d Di
\end{align*}}
 \\
 \hline
    14 & \cite{Lopez2020ModifiedSEIR} & \vtop{
 \hbox{\strut S, E, I, Q, }
 \hbox{\strut R, D, C}
 } & 
 \vtop{
 \hbox{\strut  $\beta$, $\alpha$, $\gamma$, }
 \hbox{\strut  $\delta$, $\tau$ }

 }
 & C, Q, D &  & k(t), $\lambda (t)$ &
\parbox{1cm}{\begin{align*}
        \dot{S}(t)&=\mu N-\alpha S-\beta S I N-\mu S\\
        \dot{E}(t)&=\beta S I N-\mu E-\gamma E\\
        \dot{I}(t)&=\gamma E-\delta I-\mu I \mu S\\
        \dot{Q}(t)&=\delta I - \lambda Q-k Q-\mu Q\\
        \dot{R}(t)&=\lambda Q -\mu S\\
        \dot{D}(t)&=kQ\\
        \dot{C}(t)&=\alpha S-\mu C - \tau C
\end{align*}}
 \\
 \hline
     61 & \cite{Hubbs2020SocialDistancing} & \vtop{
 \hbox{\strut S, E, I, R }
 } & 
p, $\alpha$, $\beta$, $\gamma$, K
 & KI &  &  &
\parbox{1cm}{\begin{align*}
        \dot{S}(t)&=-p \beta S I\\
        \dot{E}(t)&=p \beta S I-\alpha E\\
        \dot{I}(t)&=\alpha E -\gamma I\\
        \dot{R}(t)&=\gamma I
\end{align*}}
 \\
 \hline
      5 & \cite{pribylova2020seiar} & \vtop{
 \hbox{\strut S, E, I,  }
 \hbox{\strut R, A}
 } & 
$\mu_1$, $\mu_2$, $\gamma$, $\beta$, p
 & I &  &  &
\parbox{1cm}{\begin{align*}
        \dot{S}(t)&=- \beta S( I+A)\\
        \dot{E}(t)&=\beta S( I+A)-\gamma E\\
        \dot{I}(t)&=\gamma p E -\mu_1 I\\
        \dot{A}(t)&=\gamma (1-p) E -\mu_2 A\\
        \dot{R}(t)&=\mu_1 I+\mu_2 A
\end{align*}}
 \\
 \hline
       1 & \cite{Wen2020OptimalStrategies} & \vtop{
 \hbox{\strut S, E, I,  }
 \hbox{\strut R, Q}
 } & 
$\phi$, $\beta$, $\gamma$, w
 & Q &  &  &
\parbox{1cm}{\begin{align*}
        \dot{S}(t)&=- \beta SI\\
        \dot{E}(t)&=\beta SI-w E\\
        \dot{I}(t)&=w E -\phi I-(1-\phi)\gamma I\\
        \dot{R}(t)&=\gamma Q+(1-\phi)\gamma I\\
        \dot{Q}(t)&=-\gamma Q+\phi I
\end{align*}}
 \\
 \hline
        3 & \cite{Liangrong2020Epidemic_Analysis} & \vtop{
 \hbox{\strut S, E, I, R }
 \hbox{\strut Q, D, P}
 } & 
\vtop{
 \hbox{\strut $\alpha$, $\beta$, $\gamma$, $\delta$ }
 \hbox{\strut $E_0$, $I_0$}
 }
 & Q, R, D &\parbox{1cm}{\begin{align*}
        & S_0\\
        & E_0\\
        & I_0\\
        & Q_0\\
        & R_0\\
        & D_0\\
        & P_0
\end{align*}}  & $\lambda (t)$, $\kappa (t)$ &
\parbox{1cm}{\begin{align*}
        \dot{S}(t)&=- \beta SI/N-\alpha S\\
        \dot{E}(t)&=\beta SI/N-\gamma E\\
        \dot{I}(t)&=\gamma E -\delta I\\
        \dot{Q}(t)&=\delta I-\lambda Q- \kappa Q\\
        \dot{R}(t)&=\lambda Q\\
        \dot{D}(t)&=\kappa Q\\
        \dot{P}(t)&=\alpha S\\
\end{align*}}
 \\
 \hline
         4 & \cite{sameni2020mathematical} & \vtop{
 \hbox{\strut S, E, I,  }
 \hbox{\strut R, P}
 } & 
\vtop{
 \hbox{\strut $\alpha_e$, $\alpha_i$, $\rho$,  }
 \hbox{\strut $\beta$, $\mu$, $\kappa$, $e_0$}
 }
 & 
  \parbox{1cm}{\begin{align*}
&(1) S+vs(t)\\
&(1) I+vi(t)\\
&(1) R+vr(t)\\
&(1) P+vp(t)\\
&(2) P
\end{align*}}
 &\parbox{1cm}{\begin{align*}
        & N-e_0\\
        & e_0\\
        & 0\\
        & 0\\
        & 0
\end{align*}}  &
\parbox{1cm}{\begin{align*}
        & vs(t)\\
        &  vi(t)\\
        & vr(t)\\
        & vp(t)
\end{align*}}
 &
\parbox{1cm}{\begin{align*}
        \dot{S}(t)&=- \alpha_e SE-\alpha_i S I\\
        \dot{E}(t)&=\alpha_e SE+\alpha_i S I-\kappa E-\rho E\\
        \dot{I}(t)&=\kappa E-\beta I-\mu I\\
        \dot{R}(t)&=\beta I +\rho E \\
        \dot{P}(t)&=\mu I
\end{align*}}
 \\
 \hline
          8 & \cite{EIKENBERRY2020Mask} & \vtop{
 \hbox{\strut S, E, I, R }
 \hbox{\strut A, H, D}
 } & 
\vtop{
 \hbox{\strut $\delta$, $\alpha$, $\eta$,   }
 \hbox{\strut $\sigma$, $\gamma_a$, $\gamma_i$, $\gamma_h$}
 \hbox{\strut $\Phi$, $E_0$, $R_0$, $A_0$}
 \hbox{\strut $H_0$, $D_0$}
 }
 & D, H, I &\parbox{1cm}{\begin{align*}
        & 1e6\\
        & E_0\\
        & 50\\
        & A_0\\
        & H_0 \\
        & R_0 \\
        & D_0
\end{align*}}  & $\beta (t)$ &
\parbox{1cm}{\begin{align*}
        \dot{S}(t)&=- \beta (I+\eta A)S/N\\
        \dot{E}(t)&=\beta (I+\eta A)S/N-\sigma E\\
        \dot{I}(t)&=\alpha \sigma E- \Phi I -\gamma_i I\\
        \dot{A}(t)&=(1-\alpha)\sigma E-\gamma_a A\\
        \dot{H}(t)&=\Phi I-\delta H-\gamma_h H\\
        \dot{R}(t)&=\gamma_i I+\gamma_a A+\gamma_h H\\
        \dot{D}(t)&=\delta H
\end{align*}}
 \\
 \hline
           38 & \cite{Shi2020SEIR_Transmission} & \vtop{
 \hbox{\strut S, E, I,}
 \hbox{\strut Sq, Eq, H, R}
 } & 
\vtop{
 \hbox{\strut c, q, $\lambda$, $\beta$ }
 \hbox{\strut $\delta_i$, $\delta_q$, $\alpha$, $\gamma_i$}
 \hbox{\strut $\gamma_h$, $\theta$}
 }
 &   \parbox{1cm}{\begin{align*}
&(1) I, R\\
&(2) H, vi I, vr R\\
&(3) Sq, Eq
\end{align*}} &  & $\sigma (t)$ &
\parbox{1cm}{\begin{align*}
        \dot{S}(t)&=- (c \beta +cq(1-\beta))S(I+\theta E)+
        \lambda Sq\\
        \dot{E}(t)&=c \beta(1-q)S(I+\theta E)-\sigma E\\
        \dot{I}(t)&=\sigma E-(\delta_i+\alpha+\gamma_i) I\\
        \dot{Sq}(t)&=cq(1-\beta)S(I+\theta E)-\lambda Sq\\
        \dot{Eq}(t)&=cq \beta S(I+\theta E)-\delta_q Eq\\
        \dot{H}(t)&=\delta_i I+ \delta_q Eq-(\alpha+\gamma_h)H\\
        \dot{R}(t)&=\gamma_i I+\gamma_h H
\end{align*}}
 \\
 \hline
            41 & \cite{roosa2019assessing} & \vtop{
 \hbox{\strut S, E, A,}
 \hbox{\strut I, J, R, C}
 } & 
\vtop{
 \hbox{\strut $\beta$, k, $\gamma_1$ }
 \hbox{\strut $\gamma_2$, $\alpha$, $\rho$, q}
 }
 &   \parbox{1cm}{\begin{align*}
&(1) C\\
&(2) J, I
\end{align*}} &  &  &
\parbox{1cm}{\begin{align*}
        \dot{S}(t)&=- \beta S (I+J+qA)/N\\
        \dot{E}(t)&=\beta S (I+J+qA)/N-kE\\
        \dot{A}(t)&=k(1-\rho)E-\gamma_1 A\\
        \dot{I}(t)&=k \rho E-(\alpha+\gamma_1)I\\
        \dot{J}(t)&=\alpha I-\gamma_2 J\\
        \dot{R}(t)&=\gamma_1 (A+I)+\gamma_2 J\\
        \dot{C}(t)&=\alpha I
\end{align*}}
 \\
 \hline
             39 & \cite{Ryan2013Github} & \vtop{
 \hbox{\strut S, E, I,}
 \hbox{\strut De, Di, R, F}
 } & 
\vtop{
 \hbox{\strut  q, $\mu_i$, $\mu_d$,  }
 \hbox{\strut $\sigma$, $\sigma_d$, $\theta_e$, $\theta_i$,}
  \hbox{\strut $\gamma$, $\gamma_d$, $\beta$, $\beta_d$, }
  \hbox{\strut $\phi_i$, $\phi_e$ }
 }
 &   \parbox{1cm}{\begin{align*}
&(1) F, Di\\
&(2) F, De, Di
\end{align*}} &  &  &
\parbox{1cm}{\begin{align*}
        \dot{S}(t)&=-\beta SI/N-q \beta_d S Di/N\\
        \dot{E}(t)&=\beta SI/N+q \beta_d S Di/N-
        \sigma E-\theta_e \phi_e E\\
        \dot{I}(t)&=\sigma E- \gamma I-\mu_i I-\theta_i \phi_i I\\
        \dot{De}(t)&=\theta_e \phi_e E-\sigma_d De\\
        \dot{Di}(t)&=\theta_i \phi_i I+\sigma_d De-
        \gamma_d Di -\mu_d Di\\
        \dot{R}(t)&=\gamma I+\gamma_d Di\\
        \dot{F}(t)&=\mu_i I+\mu_d Di
\end{align*}}
 \\
 \hline
              17 & \cite{CHATTERJEE2020Healthcare} & \vtop{
 \hbox{\strut S, E, I,}
 \hbox{\strut R, Q, D}
 } & 
\vtop{
 \hbox{\strut $\beta$, $\epsilon$, $\gamma$, d, }
 \hbox{\strut q, qt}
 }
 &   \parbox{1cm}{\begin{align*}
&(1) D\\
&(2) Q, D\\
&(3) D, I, Q
\end{align*}} & 
 \parbox{1cm}{\begin{align*}
        &249 \\
        & E_0 \\
        & q Q_0\\
        & Q_0 \\
        & 23 \\
        & 5
\end{align*}}
 &  &
\parbox{1cm}{\begin{align*}
        \dot{S}(t)&=-\beta SI\\
        \dot{E}(t)&=\beta SI-\epsilon E\\
        \dot{I}(t)&=\epsilon E -\gamma I - dI -qI\\
        \dot{Q}(t)&=qI-qt Q-dQ\\
        \dot{R}(t)&=\gamma I+qt Q\\
        \dot{D}(t)&=dI+dQ
\end{align*}}
 \\
 \hline
               18 & \cite{Jia2020Modeling_control} & \vtop{
 \hbox{\strut S, E, I, A, }
 \hbox{\strut R, Q, D}
 } & 
\vtop{
 \hbox{\strut $\beta$, $\theta$, $\lambda$, $\sigma$, }
 \hbox{\strut $\rho$, $\epsilon_a$, $\epsilon_i$, $\gamma_i$,}
 \hbox{\strut $\gamma_a$, $\gamma_d$, p, $d_i$, $d_d$}
 }
 &  \parbox{1cm}{\begin{align*}
&(1) D, I, Q\\
&(2) Q, D\\
&(3) D
\end{align*}} & 
 \parbox{1cm}{\begin{align*}
        &9219849e3\\
        & 4142251e2 \\
        & 3207\\
        & 595 \\
        & 563 \\
        & 227 \\
        & 3
\end{align*}}
 &  &
\parbox{1cm}{\begin{align*}
        \dot{S}(t)&=-\beta S(I+\theta A)-pS+\lambda Q\\
        \dot{Q}(t)&=pS-\lambda Q\\
        \dot{E}(t)&=\beta S(I+\theta A)-\sigma E\\
        \dot{A}(t)&=\sigma (1-\rho)E-\epsilon_a A-\gamma_a A\\
        \dot{I}(t)&=\sigma \rho E-\gamma_i I -d_i I-\epsilon_i I \\
        \dot{D}(t)&=\epsilon_a A+\epsilon_i I -d_d D-\gamma_d D\\
        \dot{R}(t)&=\gamma_a A+\gamma_i I+\gamma_d D
\end{align*}}
 \\
 \hline
     31 & \cite{Dohare2020Mathematical_Model} & \vtop{
 \hbox{\strut S, E, I, A, }
 \hbox{\strut J, R}
 } & 
\vtop{
 \hbox{\strut $\alpha$, $\sigma$, h, }
 \hbox{\strut r, q, f, $\beta_1$,}
 \hbox{\strut $\beta_2$, $\phi$, $\gamma$, $I_0$}
 }
 &  \parbox{1cm}{\begin{align*}
&(1) I, J\\
&(2) I
\end{align*}} & 
 \parbox{1cm}{\begin{align*}
        &0.9 N\\
        & 9(I_0+A_0) \\
        & I_0\\
        & I_0 f \\
        & 0 \\
        & 0 
\end{align*}}
 &  &
\parbox{1cm}{\begin{align*}
        \dot{S}(t)&=-\alpha \dfrac{E+I+A}{N}S-\sigma S\\
        \dot{E}(t)&=\alpha \dfrac{E+I+A}{N}S-\beta_1 E\\
        \dot{I}(t)&=\beta_1 h E +\beta_2 r A-\phi q I-
        \gamma (1-q)I\\
        \dot{A}(t)&=\beta_1 (1-h)E-\beta_2 rA-
        \gamma(1-r)A\\
        \dot{J}(t)&=\phi q I-\gamma J \\
        \dot{R}(t)&=\gamma(1-q)I+\gamma(1-r)A+\gamma J
\end{align*}}
 \\
 \hline
 32 & \cite{Rahman2020Modelling_transmission} & S, L, I, R & 
\vtop{
 \hbox{\strut  w, $\beta$, $\alpha$,  }
 \hbox{\strut $\gamma$, $\mu$}
 }
 & wL &  &  &
\parbox{1cm}{\begin{align*}
        \dot{S}(t)&=A- \dfrac{\beta S I}{1+\alpha I}-\mu S\\
        \dot{L}(t)&=\dfrac{\beta S I}{1+\alpha I}-(w+\mu)L\\
        \dot{I}(t)&=wL-(\gamma+\mu)I\\
        \dot{R}(t)&=\gamma I-\mu R
\end{align*}}
 \\
 \hline
                  33 & \cite{Fosu2020Construction} & \vtop{
 \hbox{\strut S, L, E, }
 \hbox{\strut I, Q, R}
 } & 
\vtop{
 \hbox{\strut $\gamma$, $\beta_1$, $\eta$, }
 \hbox{\strut  $\delta$, $\xi$, $\theta_2$,}
  \hbox{\strut $\epsilon$, $\theta_1$, $\alpha_1$, $\alpha_2$,}
 }
 & L, Q &  &  &
\parbox{1cm}{\begin{align*}
        \dot{S}(t)&=\mu N-\beta_1 S I-(\gamma+\eta)S+
        \delta L+\xi E\\
        \dot{L}(t)&=\eta S-(\gamma+\delta)L\\
        \dot{E}(t)&=\beta_1 S I-(\gamma +\theta_2+\epsilon+\xi)E\\
        \dot{I}(t)&=\epsilon E-(\gamma +\theta_1+\alpha_2)I\\
        \dot{Q}(t)&=\theta_1 I+\theta_2 E-(\gamma+\alpha_2)Q\\
        \dot{R}(t)&=\alpha_1 I+\alpha_2 Q -\gamma R
\end{align*}}
 \\
 \hline
 \caption{List of SEIR models and their main features}
 \label{List_SEIR_models}
 \end{longtable}
} 

\section{Results and Discussion}\label{sec:results}


We analysed the structural identifiability and observability of the 17 SIR model structures (a total of 98 model versions considering the different output configurations and time-varying parameter assumptions) and 19 SEIR models (with a total of 157 model versions) listed in Tables \ref{List_SIR_models} and \ref{List_SEIR_models}. 
The detailed results for each model are given in \ref{sec:appendix}, which reports the structural identifiability of each parameter and the observability of each state, for every model version. 
In the remainder of this section we provide an overview of the main results.

\subsection{General patterns}

The general patterns regarding state observability are as follows. The recovered state (R) is almost never observable unless it is directly measured (D.M.) as output; the only exceptions are two SEIR models, 31 and 38, for which R is observable under the assumption of time-varying parameters. The susceptible state (S), in contrast, is observable in roughly two thirds of the models (SIR: 65/98, SEIR: 103/157); this is also true for the exposed state (E) in the SEIR models.
The infected state (I) is included in most studies among the outputs, either directly (D.M.) or indirectly measured (as part of a parameterized measurement function). When it is not considered in this way, its observability is generally similar to that of S (in 18/157 model versions I is not an output and it is observable in 13/18). 

The transmission and recovery rates ($\beta$, $\gamma$) are the two parameters common to all SIR models. The transmission rate is identifiable in 59/98 model versions, and $\gamma$ in 51/98 and its derivatives in 12/98.
SEIR models have a third parameter in common, the latent period ($\kappa$). It is identifiable in most of the models (145/157), as well as the recovery rate (111/157). The transmission rate is identifiable in 101/157 model versions, but it is not identifiable in any SEIR model version that accounts for social distancing (numbers 34 and 61); we found no clear pattern in the other models. 

\subsection{The effect of time-varying parameters}

\begin{figure}[H]
\centering\includegraphics[width=\linewidth]{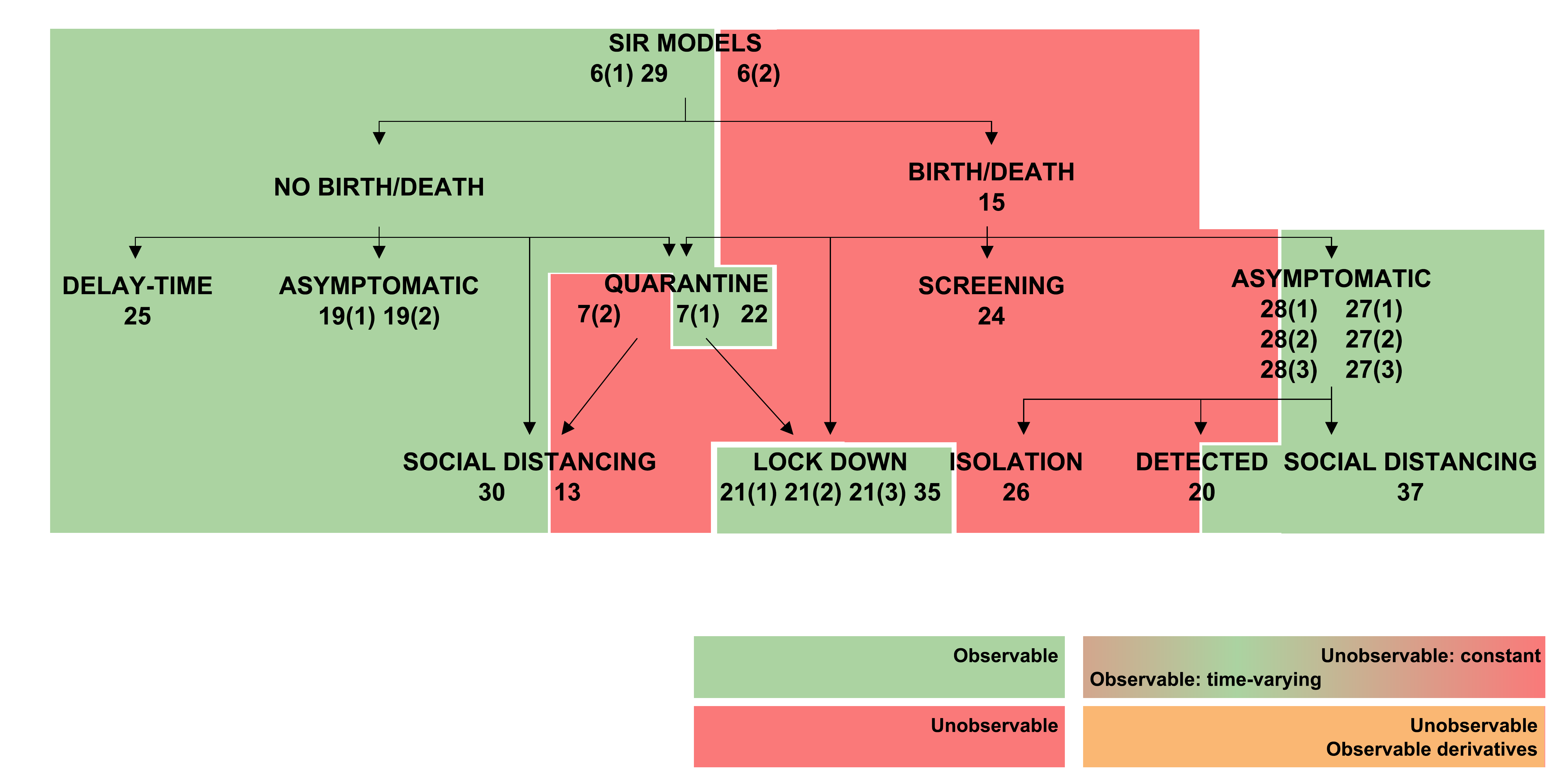}
\caption{Observability of $\beta$ (transmission rate) in SIR models. Models in which $\beta$ is observable are shown in green, and non-observable in red.}
\label{SIR_beta}

\end{figure}

\begin{figure}[H]
\centering\includegraphics[width=\linewidth]{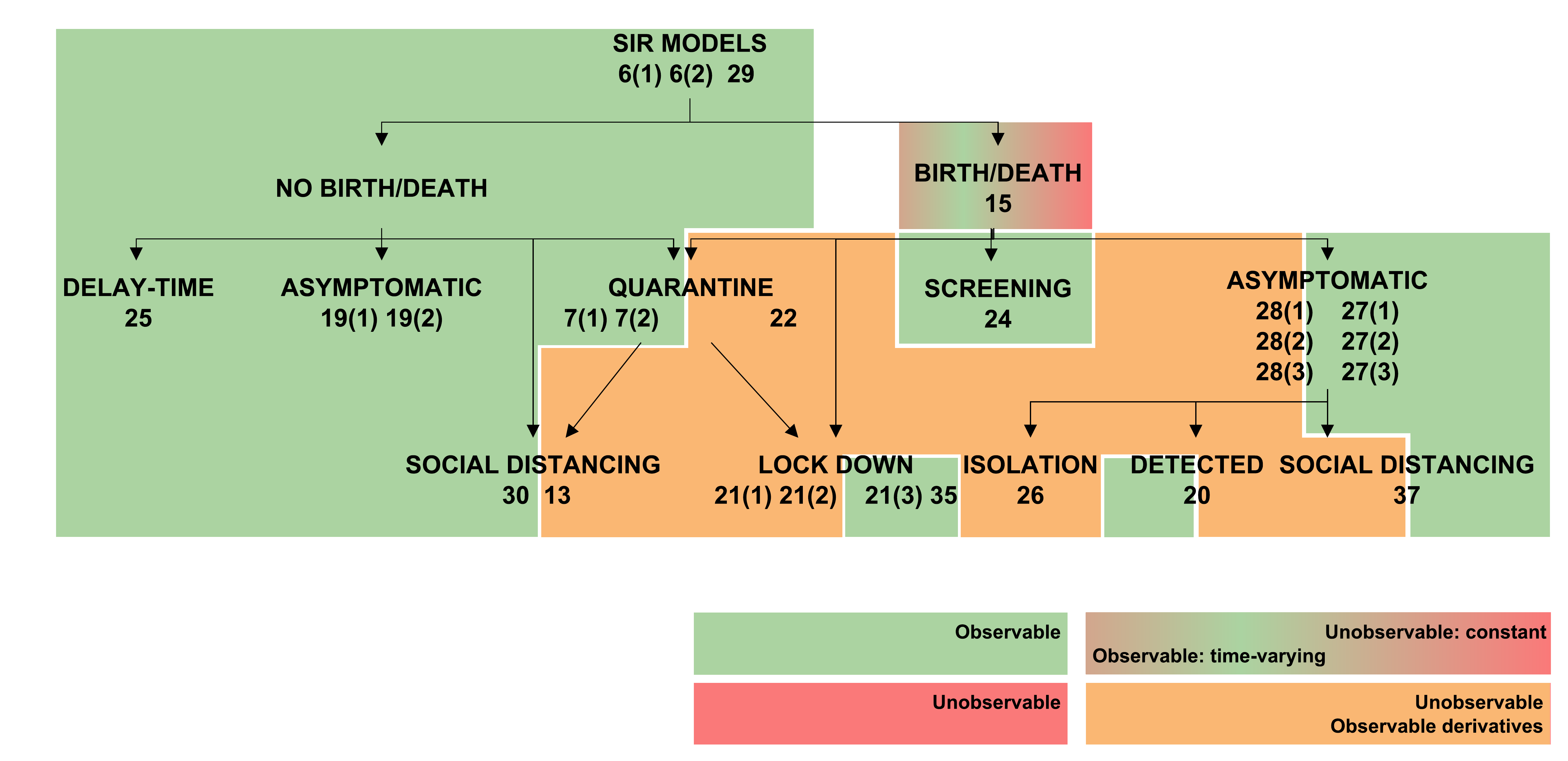}
\caption{Observability of $\gamma$ (recovery rate) in SIR models. Models in which $\gamma$ is observable are shown in green. Models in which it is unobservable if constant and observable if time-varying are shown in a green-red gradient. Finally, models in which only its time-derivatives are observable are shown in orange.}
\label{SIR_gamma}
\end{figure}

\begin{figure}[H]
\centering\includegraphics[width=1\linewidth]{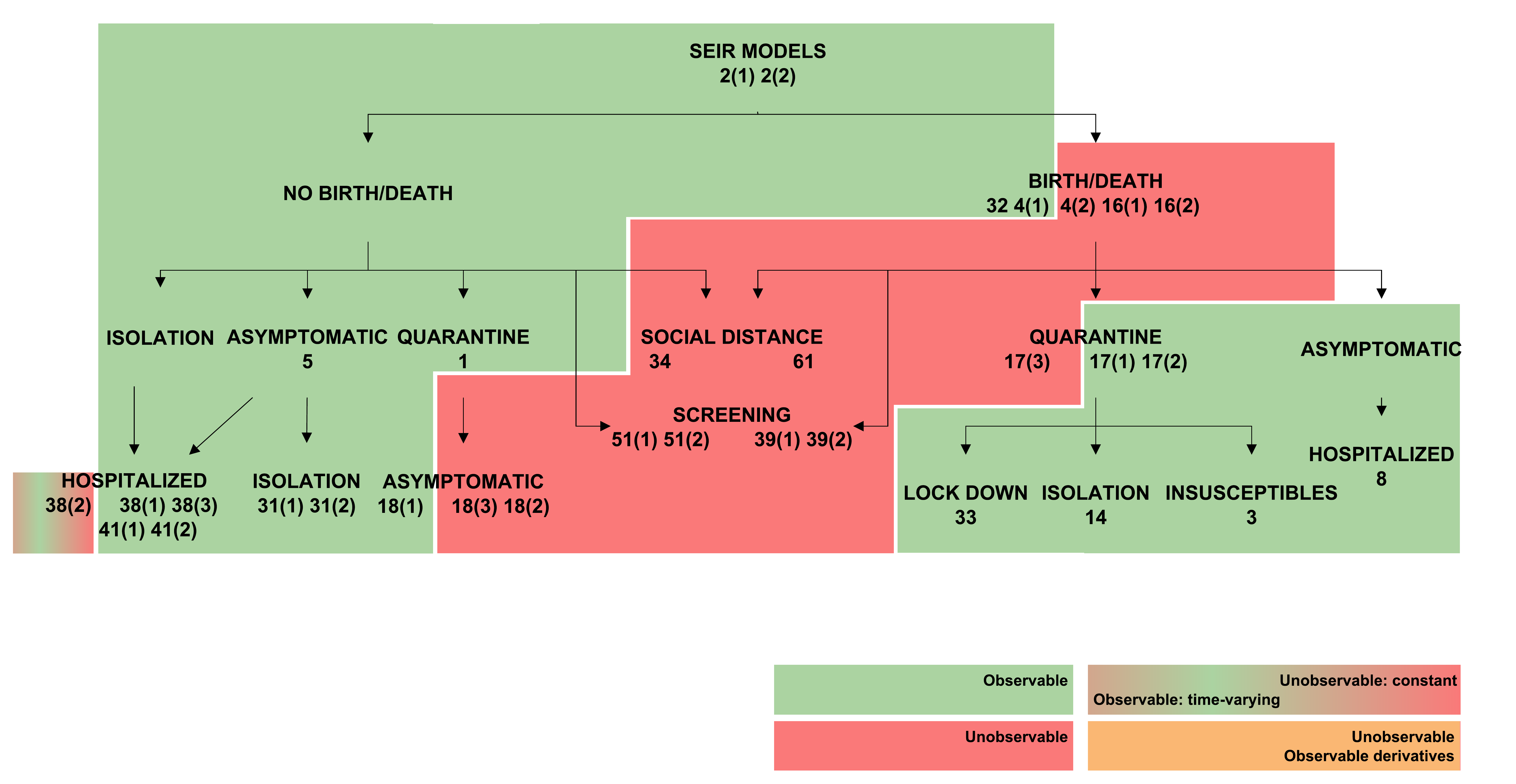}
\caption{Observability of $\beta$ (transmission rate) in SEIR models. Models in which $\beta$ is observable are shown in green, and non-observable in red. Models in which it is unobservable if constant and observable if time-varying are shown in a green-red gradient.}
\label{SEIR_beta}
\end{figure}

\begin{figure}[H]
\centering\includegraphics[width=1\linewidth]{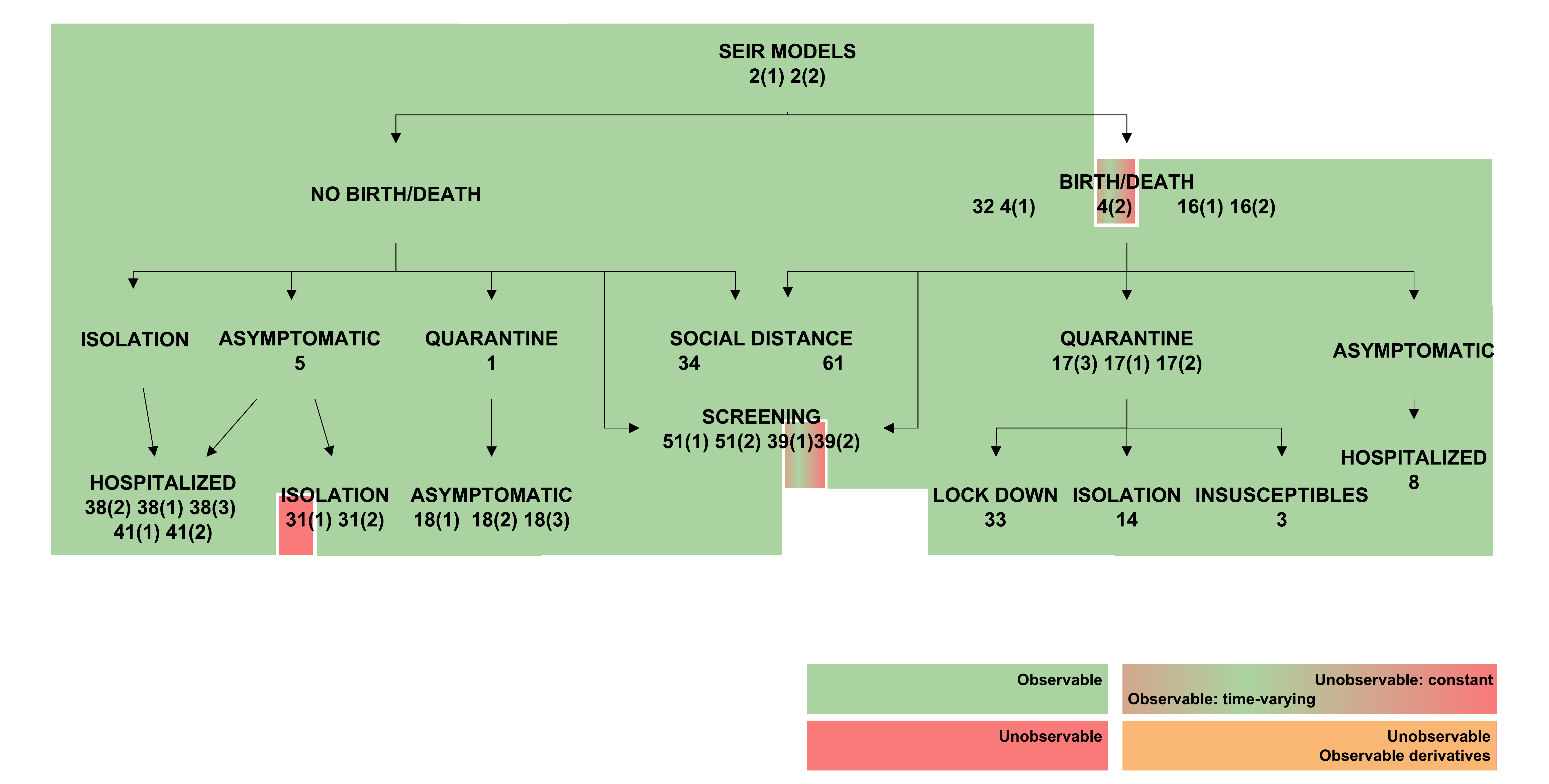}
\caption{Observability of $\kappa$ (latent period) in SEIR models. Models in which $\kappa$ is observable are shown in green, and non-observable in red. Models in which it is unobservable if constant and observable if time-varying are shown in a green-red gradient.}
\label{SEIR_k}
\end{figure}

\begin{figure}[H]
\centering\includegraphics[width=1\linewidth]{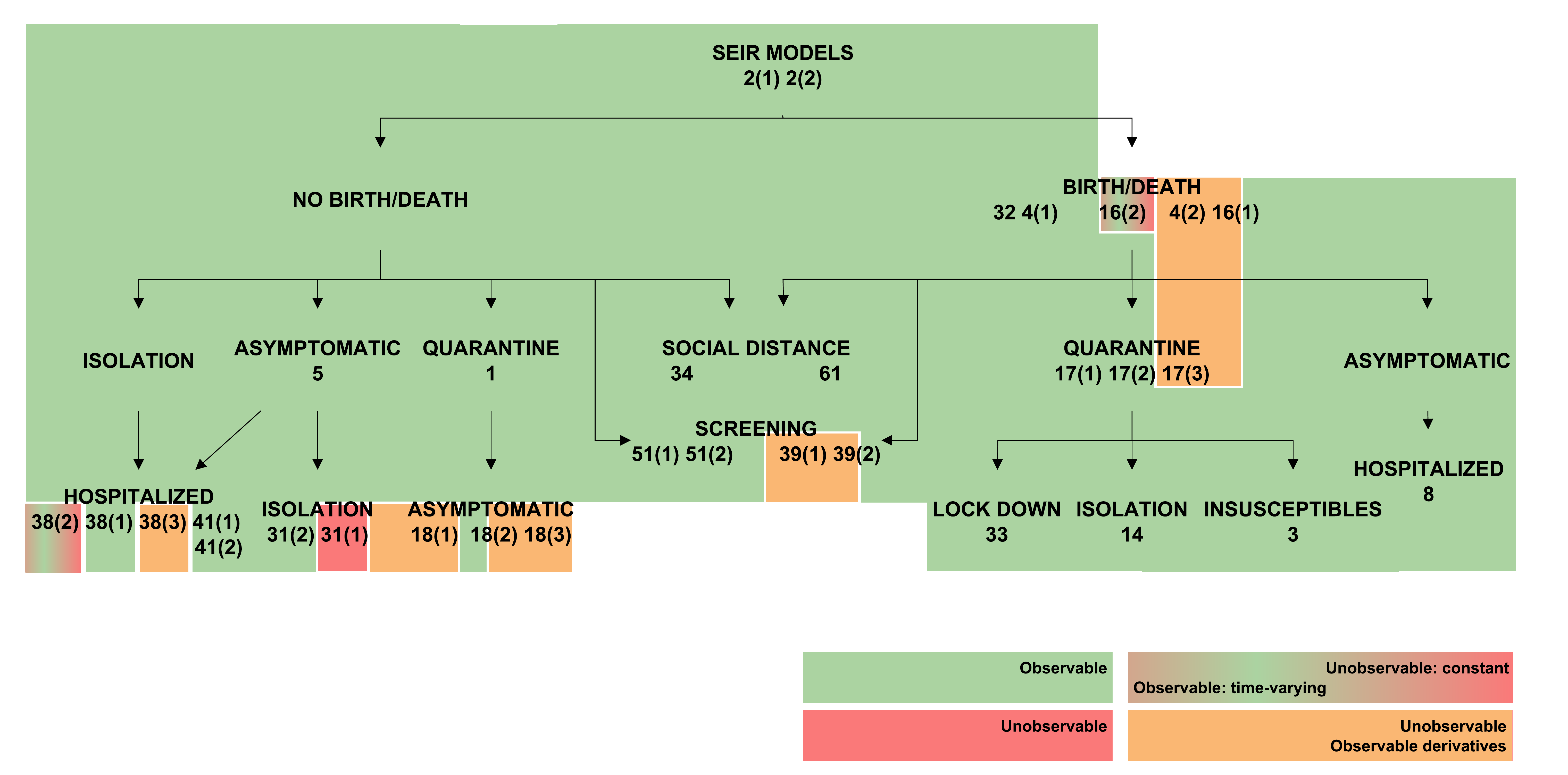}
\caption{Observability of $\gamma$ (recovery rate) in SEIR models. Models in which $\gamma$ is observable are shown in green, and non-observable in red. Models in which it is unobservable if constant and observable if time-varying are shown in a green-red gradient. Finally, models in which only its time-derivatives are observable are shown in orange.}
\label{SEIR_gamma}
\end{figure}

The transmission rate $\beta$, the recovery rate $\gamma$, and in SEIR models the latent period $\kappa$, can vary during an epidemic as a result of changes in the population's behaviour \cite{Liangrong2020Epidemic_Analysis,chen2020timedependent}, the introduction of new drugs or new medical equipment \cite{Liangrong2020Epidemic_Analysis}, or the reduction of the period duration as a result of high temperatures \cite{Fuxiang2019PeriodicSEIRS}. 
To account for such variations, the present study has considered both the constant and the time-varying cases, by including the corresponding variables either in the constant parameter vector $\theta$ or in the unknown input vector $w(t)$, respectively, as described in Remark \ref{remark_time}.
Changing a parameter from constant to time-varying naturally influences structural identifiability and observability. This effect is graphically summarized in Figures \ref{SIR_beta}-\ref{SEIR_gamma}, which represent classes of models in tree form and classify them according to their observability. Each model is shaded with a color, according to the observability of the parameter studied. Some models include different rates for different population groups: for example, they may consider two different transmission rates for symptomatic and asymptomatic individuals. For those models, each rate may have different observability properties when considered time-varying parameters; in such cases the model is depicted between two color blocks (see for example the SIR 20 model in Figure \ref{SIR_beta}). 

Changing $\beta$ from a constant to a time-varying parameter (or equivalently an unknown input) does not change its observability nor that of the other variables in SIR models. In contrast, this is not the case with the recovery rate $\gamma$, for which a somewhat counter-intuitive result may be obtained: by changing $\gamma$ from a constant to a continuous function of time with at least a non-zero derivative, its model can become more observable and identifiable -- despite the fact that it is an unknown function. An example of this is the SIR model 15: if $\gamma$ is constant the model has only one identifiable parameter, $\tau$, and no observable states; if $\gamma$ is time-varying with at least one non-zero derivative, two parameters become identifiable ($\beta$, $\mu$), two states become observable (I, S), and $\gamma$ itself is observable. In the other models, when $\gamma$ is not identifiable as a constant nor observable as an unknown input, its successive derivatives are observable. 


For the SEIR models, the consideration of the $\beta$ parameter as an unknown input function follows a similar trend to that of the SIR models with the exception of model 38, which gains both observability and identifiability and becomes FISPO. 
Considering the recovery rate $\gamma$ (Fig. \ref{SEIR_gamma}) or the latent period $\kappa$ (Fig. \ref{SEIR_k}) individually as time-varying parameters generally leads to greater observability, except for model 31(1). As an example, in model 39(2) one of the unknown inputs becomes observable, three states become observable (S, E, I), and three parameters become identifiable ($\gamma$, $\mu_i$, $\beta$); or the 16(2) model, in which both its input and three states (S, E, I) become observable and two parameters ($\mu$, $\beta$) become identifiable.

Besides the transmission rate, latent period, and recovery rate, other rates (screening, disease-related deaths, and isolation) have also been considered as time-varying parameters in some studies. The observability of most models is not modified if these parameters are allowed to change in time; the exception being 8 models which gain observability. An example is the SEIR model 41(1), which has seven parameters, seven states, and one output. Assuming constant parameters, five of them  are structurally identifiable ($\kappa$, $\alpha$, $\beta$, $\gamma_1$, $\gamma_2$) and two are unidentifiable (q, $\rho$), while there are three observable states (I, J, C) and four unobservable states (S, E, A, R) \cite{roosa2019assessing}. However, when the parameter $\rho$ (which describes the proportion of exposed/latent individuals who become clinically infectious) is considered time-varying, all parameters become identifiable (including $\rho$) and six states become observable (all except R, which is never observable unless it can be directly measured, as we have already mentioned). 

The fact that allowing an unknown quantity to change in time can improve its observability -- and also the observability of other variables in a model -- may seem paradoxical. An intuitive explanation can be obtained from the study of the symmetry in the model structure. The existence of Lie symmetries amounts to the possibility of transforming parameters and state variables while leaving the output unchanged, i.e. their existence amounts to lack of structural identifiability and/or observability \cite{yates2009structural}. The STRIKE-GOLDD toolbox used in this paper includes procedures for finding Lie symmetries \cite{Massonis2020FindingAB}. Let us use the SIR 15 model as an example. This model has five parameters ($\tau$, $\beta$, $\rho$, $\mu$, d), of which only $\tau$ is identifiable if assumed constant. The model contains the following symmetry:
$$\mu^{*} = \mu +\epsilon \rho, $$
where $\epsilon$ is the parameter of the Lie group of transformations. Thus, there is a symmetry between $\rho$ and $\mu$ that makes them unidentifiable: changes in one parameter can be compensated by changes in the other one. However, if $\rho$ is time-varying and $\mu$ is constant, the latter cannot compensate the changes of the former, and the symmetry is broken. Indeed, if $\rho$ is considered time-varying the model gains identifiability (not only $\mu$, but also $\tau$ and $\beta$ become identifiable) and observability (S, I and $\rho$ become observable).

\subsection{Applying the results in practice: an illustrative example}
Let us now illustrate how the results of this study may be applied in a realistic scenario. We use as an example the model SIR 26, which has 6 states (S, I, R, A, Q, J) and 16 parameters ($d_1$, $d_2$, $d_3$, $d_4$, $d_5$, $d_6$, $k_1$, $k_2$, $\lambda$, $\gamma_1$, $\gamma_2$, $\epsilon_a$, $\epsilon_q$, $\epsilon_j$, $\mu_1$, $\mu_2$); its equations are shown in Table \ref{List_SIR_models}. This model includes the following additional features with respect to the basic SIR model: birth/death, asymptomatic individuals (A), quarantine (Q), and isolation (J). In its original publication two states were measured (Q, J). With these two states as outputs the model has five identifiable parameters ($d_1$, $d_5$, $\epsilon_q$, $k_2$, $\mu_1$) and two observable states (A, I); thus, there are two unobservable states (S, R) and ten unidentifiable parameters. 

If we are interested in estimating e.g. the number of susceptible individuals (S), this model would not be appropriate. How should we proceed in that scenario?

One way of improving observability could be by \textit{including more outputs (option 1)}. For example, since there is a separate class for asymptomatic individuals (A), the infected compartment (I) considers only individuals with symptoms, and we could assume that they can be detected. By including `I' in the output set, the structural identifiability and observability of the model improves: six more parameters are identifiable ($\lambda$, $\epsilon_a$, $\epsilon_j$, $d_4$, $k_1$, $\mu_2$) and the state in which we are interested (S) becomes observable. 

However, including more outputs is not always realistic. Another possibility would then be to reduce the complexity of the model by \textit{decreasing the number of additional features (option 2)}. For example, leaving out the asymptomatic compartment leads to the following model:
\begin{align*}
    \dot{S(t)}&= bN-S(I \lambda +\lambda \epsilon_q Q+\lambda \epsilon_j J+d_1)\\
    \dot{I(t)}&=S(I \lambda +\lambda \epsilon_q Q+\lambda \epsilon_j J)-(\gamma_1+\mu_1+\mu_2+d_2)I\\
    \dot{R(t)}&=\gamma_1 I+\gamma_2 J-d_3 R\\
    \dot{Q(t)}&=\mu_1 I-(k_2+d_5)Q\\
    \dot{J(t)}&=k_2 Q+\mu_2 I-(\gamma_2+d_6)J
\end{align*}
The output of the model is the same, Q, J. In this case, the model has eight identifiable parameters ($\lambda$, $\epsilon_q$, $\epsilon_j$, $d_1$, $d_5$, $\mu_1$, $\mu_2$,  $k_2$) and two observable states (S, I). 

A third possibility is to \textit{simplify the parametrization of the model (option 3)}. This model considers a different death rate for every compartment ($d_i, \ i=1,\dots,6.$). With some loss of generality, we could consider a specific death rate for infected individuals, $d_I=d_2$, and a general death rate $d$ for all non-infected and asymptomatic individuals, $d=d_1=d_3=d_4=d_5=d_6$. This reduction of the number of parameters leads to a better observability to the model: the only unidentifiable parameters are $d_2,\gamma_1,$ and $k_1,$ and the only non-observable state is R. Thus, this option also allows to identify S.


\section{Conclusions}\label{sec:conclusions}

Our analyses have shown that a fraction of the models found in the literature have unidentifiable parameters. Key parameters such as the transmission rate ($\beta$), the recovery rate ($\gamma$), and the latent period ($\kappa$) are structurally identifiable in most, but not all, models. The transmission and recovery rates are identifiable in roughly two thirds of the models, and the latent period in almost all ($>90\%$) of them.
Likewise, the states corresponding to the number of susceptible (S) and exposed (E) individuals are non-observable in roughly one third of the model versions analysed in this paper. The number of infected individuals (I) can usually be directly measured, but it is non-observable in one third of the model versions in which it is not measured.
The situation is worse for the number of recovered individuals (R), which is almost never observable unless it is directly measured.
Many models include other states in addition to S, E, I, and R, which are not always observable either. 

The transmission rate and other parameters may vary during the course of an epidemic, as a result of a number of factors such as changes in public policy, population behaviour, or environmental conditions. 
To account for these variations, in the present study we have considered both the constant and the time-varying parameter case. Somewhat unexpectedly, we found that allowing for variability in an unknown parameter often improves the observability and/or identifiability of the model. This phenomenon might be explained by the contribution of this variability to the removal of symmetries in the model structure. 

Structural identifiability and observability depend on which states or functions are measured. The lack of these properties may in principle be surmounted by choosing the right set of outputs \cite{anguelova2012minimal}, but the required measurements are not always possible to perform in practice. Epidemiological models are a clear example of this; limitations such as lack of testing or the existence of asymptomatic individuals usually make it impossible to have measurements of all states. 
An alternative to measuring more states is to use a model with fewer compartments and/or a simpler parameterization, thus decreasing the number of states and/or parameters. Reducing the model dimension in this way may achieve observability and identifiability.

Even when it is not possible (or practical) to avoid non-observability or non-identifiability by any means, the model may still be useful, as long as it is only used to infer its observable states or identifiable parameters. For example, we may be interested in determining the transmission rate $\beta$ but not the number of recovered individuals R; in such case it is fine to use a model in which $\beta$ is identifiable even if R is not observable. Of course, this means that, to ensure that a model is properly used, it is necessary to characterize its identifiability and observability in detail, to know if the quantity of interest is observable/identifiable.

The contribution of this work has been to provide such a detailed analysis of the structural identifiability and observability of a large set of compartmental models of COVID-19 presented in the recent literature. The results of our analyses can be used to avoid the pitfalls caused by non-identifiability and non-observability.
By classifying the existing models according to these properties, and arranging them in a structured way as a function of the compartments that they include, our study has answered the following question: given the sets of existing models and available measurements, which model is appropriate for inferring the value of some particular parameters, and/or to predict the time course of the states of interest?


\appendix
\clearpage
\section{Detailed structural identifiability and observability results}\label{sec:appendix}
The tables included in the following pages report the results of the observability and structural identifiability analyses of all the model variants considered in this paper. 
Each block of rows represents one of the following assumptions: 
\begin{itemize}
    \item All parameters considered constant (i.e. as is usually the case in the original publications).
    \item Transmission rate $\beta$ considered time-varying.
    \item Latent period $\kappa$ considered time-varying (only in SEIR models; SIR models do not have this parameter).
    \item Recovery rate $\gamma$ considered time-varying.
    \item All parameters considered time-varying.
\end{itemize}
Within each block, each row provides detailed information about identifiable and non-identifiable parameters, observable and non-observable states, directly measured (D.M.) states, observable and unobservable unknown inputs (and time-varying parameters), known inputs, and number of derivatives of the unknown inputs (and time-varying parameters) assumed to be non-zero (nnDerW).
The suffix $\_$\textit{d number} represents the $n^{th}$ derivative of an unknown function (e.g. $\beta\_d1$ is the first derivative of the time-varying parameter $\beta$). 

The blank blocks in the tables of the SEIR models numbers 38 and 8 indicate that the corresponding time-varying case is already considered in the original formulation of the model. The SIR models 29 and 30 have only been studied in their original form, i.e. without considering time-varying parameters, because these models do not contain the common parameters of the SIR models; instead they use the $R_0$ constant. 



\section*{Supplementary material}

## Table 1

| | | 6 | | 7 | | 13 | 20 |
|---|---|---|---|---|---|---|---|
| | | h=I | h=KI | h=I, R, Q | h=Q | h=X | h=D, R, T |
| **Constant parameters** | Identifiable | β, γ | γ | FISPO | | k0 | β, δ, η, ξ, ν, ρ |
| | Non Identifiable | | κ, β | | β, γ, δ | k, β, α | α, γ, ε, C, λ, σ, κ, θ, μ, τ |
| | Observable states | S | | | | | |
| | Non Observable states | R | R, S, I | | I, R, S | I, R, S | A, E, H, I, S |
| | D. M. | I | | | Q | X | D, R, T |
| | Observable unknown inputs | | | | | | |
| | Non Observable unknown inputs | | | | | | |
| | Known inputs | | | | | | |
| **Transmission rate as unknown input** | Identifiable | All | γ | | | k0 | ξ, ν, ρ, η |
| | Non Identifiable | | K | | γ, δ | β, k | ε, C, λ, σ, κ, θ, μ, τ |
| | Observable states | S | | | | | |
| | Non Observable states | R | R, S, I | | I, R, S | I, R, S | A, E, H, I, S |
| | D. M. | I | | | Q | X | D, R, T |
| | Observable unknown inputs | β | | | | | β, δ |
| | Non Observable unknown inputs | | β | | β | α | α, γ |
| | Known inputs | | | | | | |
| | nnDerW | 1--2 | 1--2 | | 1--2 | 1--2 | 1 |
| **Recovery rate as unknown input** | Identifiable | All | | | β, δ | k0 | β, δ, η, ν |
| | Non Identifiable | | K, β | | | α, k | α, γ, ε, C, θ, μ, τ |
| | Observable states | S | | | | | |
| | Non Observable states | R | R, S, I | | I, R, S | I, R, S | A, E, H, I, S |
| | D. M. | I | | | Q | X | D, R, T |
| | Observable unknown inputs | γ | γ | | γ_d1, γ_d2 | β_d1, β_d2 | ρ, ξ, κ_d1, λ_d1, σ_d1 |
| | Non Observable unknown inputs | | | | γ | β | κ, λ, σ |
| | Known inputs | | | | | | |
| | nnDerW | 1--2 | 1--2 | | 1--2 | 1--2 | 1 |
| **Unknown inputs** | Identifiable | | | | | | |
| | Non Identifiable | | | | | | |
| | Observable states | S | | | | | |
| | Non Observable states | R | R, S, I | | I, R, S | I, R, S | A, E, H, I, S |
| | D. M. | I | | | Q | X | D, R, T |
| | Observable unknown inputs | β, γ | γ | | | k0 | β, δ, η, ξ, ν, ρ |
| | Non Observable unknown inputs | | K, β | | β, γ, δ | k, β, α | α, γ, ε, C, λ, σ, κ, θ, μ, τ |
| | Known inputs | | | | | | |
| | nnDerW | 1 | 1 | | 1 | 1 | 1 |

## Table 2

| | | 25 | 21 | | | 27 | | |
|---|---|---|---|---|---|---|---|---|
| | | h=A | h=I | h=C | h=C, D | h=I, R | h=R, D | h=I, R, D |
| **Constant parameters** | Identifiable | β, σ | β, p, q | β, p, q | β, μ, p, q, r | β, σ, γ, δ, k | FISPO | FISPO |
| | Non Identifiable | ρ, θ | r, μ | r, μ | | | | |
| | Observable states | y, z | C, S | S, I | S, I | A, S | | |
| | Non Observable states | | D, R | D, R | R | D | | |
| | D. M. | A | I | C | C, D | I, R | | |
| | Observable unknown inputs | | | | | | | |
| | Non Observable unknown inputs | z_d | | | | | | |
| | Known inputs | | | | | g | | |
| **Transmission rate as unknown input** | Identifiable | σ | p, q | p, q | All | All | | |
| | Non Identifiable | ρ, θ | | r, μ | | | | |
| | Observable states | y, z | C, S | S, I | S, I | A, S | | |
| | Non Observable states | | D, R | D, R | R | D | | |
| | D. M. | A | I | C | C, D | I, R | | |
| | Observable unknown inputs | β | β | β | β | β, σ | | |
| | Non Observable unknown inputs | z_d | | | | | | |
| | Known inputs | | | | | | | |
| | nnDerW | 1--2 | 1 | 1--2 | 1--2 | 1 | | |
| **Recovery rate as unknown input** | Identifiable | β | p, q, β | p, q, β | All | All | | |
| | Non Identifiable | ρ, θ | μ | μ | | | | |
| | Observable states | y, z | C,S | S, I | S, I | A, S | | |
| | Non Observable states | | D, R | D, R | R | D | | |
| | D. M. | A | I | C | C, D | I, R | | |
| | Observable unknown inputs | σ | r_d1, r_d2 | r_d1, r_d2 | r | γ | | |
| | Non Observable unknown inputs | z_d | r | r | | | | |
| | Known inputs | | | | | | | |
| | nnDerW | 1--2 | 1--2 | 1--2 | 1--2 | 1--2 | | |
| **Unknown inputs** | Identifiable | | | | | | | |
| | Non Identifiable | | | | | | | |
| | Observable states | y, z | C, S | S, I | S, I | A, S | | |
| | Non Observable states | | D, R | D, R | R | D | | |
| | D. M. | A | I | C | C, D | I, R | | |
| | Observable unknown inputs | β, σ | β, p, q | β, p, q | β, μ, p, q, r | β, σ, γ, δ, k | | |
| | Non Observable unknown inputs | ρ, θ, z_d | r, μ | r, μ | | | | |
| | Known inputs | | | | | g | | |
| | nnDerW | 1 | 1 | 1 | 1 | 1 | | |

| | | 37 | 28 | | 35 | 24 |
|---|---|---|---|---|---|---|
| | | h=Sd, I | h=I, R | h=R, D | h=I, R, D | h=Q, L | h=KI |
| **Constant parameters** | Identifiable | f, h2, βa, βi, γai, εa, εs | γ, δ, β | γ, δ, β | γ, δ, β | γ, δ, β, θi, α1, α2, η, μ | K, μ, γ |
| | Non Identifiable | γir, δ | k, v, σ | k, v, σ | k, v, σ | | φ, c |
| | Observable states | Sn, Ad, An | | I | | I, S | S, I |
| | Non Observable states | R | A, RR, S, D | A, RR, S | A, RR, S | R | R |
| | D. M. | Sd, I | I, R | R, D | I, R, D | Q, L | |
| | Observable unknown inputs | | | | | | σ |
| | Non Observable unknown inputs | | | | | | |
| | Known inputs | | g | g | g | | |
| **Transmission rate as unknown input** | Identifiable | f, h2, γai, εa, εs | δ, γ | δ, γ | δ, γ | All | K, μ, γ |
| | Non Identifiable | γir, δ | k, v | k, v | k, v | | |
| | Observable states | Sn, Ad, An | | I | | I, S | S, I |
| | Non Observable states | R | A, RR, D, S | A, RR, S | A, RR, S | R | R |
| | D. M. | Sd, I | I, R | R, D | I, R, D | Q, L | |
| | Observable unknown inputs | βa, βi | β | β | β | β | σ |
| | Non Observable unknown inputs | | σ | σ | σ | | c, φ |
| | Known inputs | | g | g | g | | |
| | nnDerW | 1 | 1--2 | 1--2 | 1--2 | 1--2 | 1 |
| **Recovery rate as unknown input** | Identifiable | f, h2, βa, βi, εa, εs | β, δ | β, δ | β, δ | All | K, μ |
| | Non Identifiable | δ | k, σ | k, σ | k, σ | | c, φ |
| | Observable states | Sn, Ad, An | | I | | S, I | S, I |
| | Non Observable states | R | A, RR, D, S | A, RR, S | A, RR, S | R | R |
| | D. M. | Sd, I | I, R | R, D | I, R, D | L, Q | |
| | Observable unknown inputs | γai, γir_d1, γir_d2 | γ, v_d1, v_d2 | γ, v_d1, v_d2 | γ, v_d1, v_d2 | α_1, α_2 | γ, σ |
| | Non Observable unknown inputs | γir | v | v | v | | |
| | Known inputs | | g | g | g | | |
| | nnDerW | 1--2 | 1--2 | 1--2 | 1--2 | 1--2 | 0--1 |
| **Unknown inputs** | Identifiable | | | | | | |
| | Non Identifiable | | | | | | |
| | Observable states | Sn, Ad, An | | I | | I, S | S, I |
| | Non Observable states | R | A, RR, S, D | A, RR, S | A, RR, S | R | R |
| | D. M. | Sd, I | I, R | R, D | I, R, D | Q, L | |
| | Observable unknown inputs | f, h2, βa, βi, γai, εa, εs | γ, δ, β | γ, δ, β | γ, δ, β | γ, δ, β, θi, α1, α2, η, μ | μ, γ, K, σ |
| | Non Observable unknown inputs | γir, δ | k, v, σ | k, v, σ | k, v, σ | | φ, c |
| | Known inputs | | g | g | g | | |
| | nnDerW | 1 | 1 | 1 | 1 | 1 | 1 |

| | | 19 | 22 | 15 | 26 | 29 | 30 |
|---|---|---|---|---|---|---|---|
| | | h=R | h=I, R | h=Q | h=τ+ρI | h=Q, J | h=I | h=I |
| **Constant parameters** | Identifiable | α, β, η, ξ | α, β, η, ξ | d, β, δ | τ | d_1, d_5, εq, k_2, μ_1 | FISPO | FISPO |
| | Non Identifiable | | | ε, γ, α_1, α_2 | β, ρ, μ, d | λ, εa, εj, k1, γ_1, γ_2, μ_2, d2, d3, d4, d6 | | |
| | Observable states | J, S, I | J, S | I, S | | A, S | | |
| | Non Observable states | U | U | R | I, R, S | I, R | | |
| | D. M. | R | I, R | Q | | Q, J | | |
| | Observable unknown inputs | | | | | | | |
| | Non Observable unknown inputs | | | | | | | |
| | Known inputs | | | | | | | |
| **Transmission rate as unknown input** | Identifiable | All | All | d, δ | τ | d_1, d_5, εq, k_2, μ_1 | | |
| | Non Identifiable | | | ε, γ, α_1, α_2 | ρ, μ, d | εa, εj, k1, γ_1, γ_2, μ_2, d2, d3, d4, d6 | | |
| | Observable states | I, J, S | J, S | I, S | | A, S | | |
| | Non Observable states | U | U | R | I, R, S | I, R | | |
| | D. M. | R | R, I | Q | | Q, J | | |
| | Observable unknown inputs | α | α | β | | | | |
| | Non Observable unknown inputs | | | | β | λ | | |
| | Known inputs | | | | | | | |
| | nnDerW | 0-1 | 1--2 | 1--2 | 1--2 | 1 | | |
| **Recovery rate as unknown input** | Identifiable | All | All | d, β, δ | τ, β, μ | d_1, d_5, εq, k_2, μ_1 | | |
| | Non Identifiable | | | α_1, α_2 | d | λ, εa, εj, k1, μ_2, d2, d3, d4, d6 | | |
| | Observable states | I, J, S | J, S | I, S | I, S | S, A | | |
| | Non Observable states | U | U | R | R | I, R | | |
| | D. M. | R | R | Q | | Q, J | | |
| | Observable unknown inputs | β, η | β, η | γ_d1, ε_d1, γ_d2, ε_d2 | ρ | γ_1_d1, γ_2_d1, γ_1_d2, γ_2_d2 | | |
| | Non Observable unknown inputs | | | γ, ε | | γ_1, γ_2 | | |
| | Known inputs | | | | | | | |
| | nnDerW | 1 | 1--2 | 1--2 | 1--4 | 1--2 | | |
| **Unknown inputs** | Identifiable | | | | | | | |
| | Non Identifiable | | | | | | | |
| | Observable states | I, J, S | J, S | I, S | | A, S | | |
| | Non Observable states | U | U | R | I, R, S | I, R | | |
| | D. M. | R | I, R | Q | | Q, J | | |
| | Observable unknown inputs | α, β, η, ξ | α, β, η, ξ | d, β, δ | τ | d_1, d_5, εq, k_2, μ_1 | | |
| | Non Observable unknown inputs | | | ε, γ, α_1, α_2 | β, ρ, μ, d | λ, εa, εj, k1, γ_1, γ_2, μ_2, d2, d3, d4, d6 | | |
| | Known inputs | | | | | | | |
| | nnDerW | 1 | 1 | 1 | 1 | 1 | | |

## Table 1

| | | 2 | 34 | 16 | | 14 | 61 |
|---|---|---|---|---|---|---|---|
| | | h=KC | h=C | h=KI | h=μ I | h=ρ I+τ | h=C, Q, D | h=KI |
| **Constant parameters** | Identifiable | γ, κ | k, β, γ | K, ε, μ, γ | ε | ε, τ | α, β, γ, δ, τ | α, γ |
| | Non Identifiable | β, K | | r, β | β, ρ, μ, d | β, ρ, μ, d | $\lambda_0, \lambda_1, k_0, k_1$ | p, β, K |
| | Observable states | | E, I, S | E, I, S | | | S, E, I | |
| | Non Observable states | S, E, I, R,C | R | R | S, E, I, R | S, E, I, R | R | S, E, I, R |
| | D. M. | | C | | | | D, Q, C | |
| | Observable unknown inputs | | | | | | | |
| | Non Observable unknown inputs | | | | | | | |
| | Known inputs | | | | | | λ, k | |
| **Transmission rate as unknown input** | Identifiable | γ, κ | All | K, ε, μ, γ | ε | ε, τ | α, γ, δ, τ | α, γ |
| | Non Identifiable | K | | r | ρ, μ, d | ρ, μ, d | $\lambda_0, \lambda_1, k_0, k_1$ | p, K |
| | Observable states | | | E, I, S | E, I, S | | E, I, S | |
| | Non Observable states | S, E, I, R, C | R | R | S, E, I, R | S, E, I, R | R | S, E, I, R |
| | D. M. | | C | | | | D, Q, C | |
| | Observable unknown inputs | β | β | | | | β | |
| | Non Observable unknown inputs | | | β | β | β | | β |
| | Known inputs | | | | | | λ, k | |
| | nnDerW | 1--2 | 1--2 | 1 | 1 | 1--2 | 1--2 | 1--2 |
| **Latent period as unknown input** | Identifiable | γ | All | K, μ, γ | | τ | α, β, δ, τ | γ |
| | Non Identifiable | β, K | | r, β | ρ, μ, d, β | β, ρ, μ, d | $\lambda_0, \lambda_1, k_0, k_1$ | p, K |
| | Observable states | | | S, E, I | E, I, S | | S, E, I | |
| | Non Observable states | S, E, I, R,C | R | R | S, E, I, R | S, E, I, R | R | S, E, I, R |
| | D. M. | | C | | | | D, Q, C | |
| | Observable unknown inputs | κ | k | ε | ε | ε | γ | α |
| | Non Observable unknown inputs | | | | | | | |
| | Known inputs | | | | | | k, λ | |
| | nnDerW | 1--2 | 1--2 | 1 | 1 | 1--2 | 1--2 | 1--2 |
| **Recovery rate as unknown input** | Identifiable | κ | All | K, μ, ε | ε | ε, τ, μ, β | α, β, γ, τ | α |
| | Non Identifiable | β, K | | r, β | μ, d, β | d | $\lambda_0, \lambda_1, k_0, k_1$ | p, K |
| | Observable states | | | S, E, I | E, I, S | S, E, I | S, E, I | |
| | Non Observable states | S, E, I, R, C | R | R | S, E, I, R | R | R | S, E, I, R |
| | D. M. | | C | | | | D, Q, C | |
| | Observable unknown inputs | γ | γ | γ | ρ_d1, ρ_d2 | ρ | δ | γ |
| | Non Observable unknown inputs | | | | ρ | | | |
| | Known inputs | | | | | | k, λ | |
| | nnDerW | 1--2 | 1--2 | 1 | 1--2 | 1--2 | 1--2 | 1--2 |
| **Unknown inputs** | Identifiable | | | | | | | |
| | Non Identifiable | | | | | | $\lambda_0, \lambda_1, k_0, k_1$ | |
| | Observable states | | | E, I, S | S, E, I | | S, E, I | |
| | Non Observable states | S, E, I, R,C | R | R | S, E, I, R | S, E, I, R | R | S, E, I, R |
| | D. M. | | C | | | | D, Q, C | |
| | Observable unknown inputs | γ, κ | k, β, γ | K, ε, μ, γ | ε | ε, τ | α, β, γ, δ, τ | γ, α |
| | Non Observable unknown inputs | β, K | | r, β | β, ρ, μ, d | β, ρ, μ, d | | p, β, K |
| | Known inputs | | | | | | λ, k | |
| | nnDerW | 1--2 | 1 | 1 | 1 | 1 | 1 | 1 |

## Table 2

| | | | 38 | | 41 | |
|---|---|---|---|---|---|---|
| | | h=I, R | h=H, vi I, vr R | h=Sq, Eq | h=C | h=J, I |
| **Constant parameters** | Identifiable | FISPO | θ, λ, δ_q | σ, c, q, β, θ, λ, δ_q | k, α, β, γ_1, γ_2 | k, α, β, γ_1, γ_2 |
| | Non Identifiable | | c, q, δ_i, β, α, vi, vr, γ_i, γ_h | γ_i, γ_h, α, δ_i | q, ρ | q, ρ |
| | Observable states | | Eq | S, E, I | I, J | I, J |
| | Non Observable states | | I, R, S, E, Sq | H, R | A, E, R, S | A, C, E, R, S |
| | D. M. | | H | Sq, Eq | C | J, I |
| | Observable unknown inputs | | σ | σ | | |
| | Non Observable unknown inputs | | | | | |
| | Known inputs | | | | | |
| **Transmission rate as unknown input** | Identifiable | | FISPO | θ, λ, δ_q, c, q | α, γ_1, γ_2, k | α, γ_1, γ_2, k |
| | Non Identifiable | | | δ_i, α, γ_h, γ_i | q, ρ | q, ρ |
| | Observable states | | | S, E, I | I, J | I, J |
| | Non Observable states | | | H, R | A, E, R, S | A, C, E, R, S |
| | D. M. | | | Sq, Eq | C | J, I |
| | Observable unknown inputs | | | β, σ | β | β |
| | Non Observable unknown inputs | | | | | |
| | Known inputs | | | | | |
| | nnDerW | | 1 | 1 | 1 | 1 |
| **Latent period as unknown input** | Identifiable | | | | α, γ_1, γ_2, β | α, γ_1, γ_2, β |
| | Non Identifiable | | | | q, ρ | q, ρ |
| | Observable states | | | | I, J | I, J |
| | Non Observable states | | | | A, E, R, S | A, C, E, R, S |
| | D. M. | | | | C | J, I |
| | Observable unknown inputs | | | | k | k |
| | Non Observable unknown inputs | | | | | |
| | Known inputs | | | | | |
| | nnDerW | | | | 1 | 1 |
| **Recovery rate as unknown input** | Identifiable | | FISPO | β, δ_q, λ, θ, c, q | β, α, k | β, α, k |
| | Non Identifiable | | | δ_i, α | q, ρ | q, ρ |
| | Observable states | | | S, E, I | I, J | I, J |
| | Non Observable states | | | H, R | A, E, R, S | A, C, E, R, S |
| | D. M. | | | Sq, Eq | C | J, I |
| | Observable unknown inputs | | | σ, γ_i_d1 | γ_1, γ_2 | γ_1, γ_2 |
| | Non Observable unknown inputs | | | γ_h, γ_i | | |
| | Known inputs | | | | | |
| | nnDerW | | 1--2 | 1--2 | 1 | 1--2 |
| **Unknown inputs** | Identifiable | | FISPO | | | |
| | Non Identifiable | | | | | |
| | Observable states | | | S, E, I | I, J, S, A, E | S, A, E |
| | Non Observable states | | | H, R | R | C, R |
| | D. M. | | | Sq, Eq | C | I, J |
| | Observable unknown inputs | | | σ, c, q, β, θ, λ, δ_q | k, α, β, γ_1, γ_2, q, ρ | k, α, β, γ_1, γ_2, q, ρ |
| | Non Observable unknown inputs | | | γ_i, γ_h, α, δ_i | | |
| | Known inputs | | | | | |
| | nnDerW | | 1--2 | 1 | 1 | 1 |

## Table 1

| | | 5 | 1 | 3 | 4 | | 8 |
|---|---|---|---|---|---|---|---|
| | | h=I | h=Q | h= Q, R, D | h=S+vs, I+vi, R+vr, P+vp | h=P | h=D, H, I |
| Constant parameters | Identifiable | ρ, β, γ, μ_1, μ_2 | v, β, γ, ψ | k, α, β, δ, γ | α_e, α_i, ρ, β, μ, κ | | δ, σ, γ_h, γ_i, γ_a, ψ |
| | Non Identifiable | | | | | α_e, α_i, ρ, β, μ, κ | α, η |
| | Observable states | A, S, E | E, I, S | S, E, I | E, I, S | | |
| | Non Observable states | R | R | P | R, P | E, I, S, R | A, E, R, S |
| | D. M. | I | Q | Q, R, D | | P | I, H, D |
| | Observable unknown inputs | | | k | vi, vs, vr_d1, vp_d1 | | β |
| | Non Observable unknown inputs | | | λ | vr, vp | | |
| | Known inputs | | | | | | |
| Transmission rate as unknown input | Identifiable | ρ, γ, μ_1, μ_2 | All | All | All | | |
| | Non Identifiable | | | | | ρ, β, μ, κ | |
| | Observable states | A, S, E | E, I, S | S, E, I | S, E, I | | |
| | Non Observable states | R | R | P | P, R | E, I, S, R | |
| | D. M. | I | Q | Q, R, D | | P | |
| | Observable unknown inputs | β | β | k, β | vi, vs, α_e, α_i, vr_d1, vp_d1 | | |
| | Non Observable unknown inputs | | | λ | vr, vp | α_e, α_i | |
| | Known inputs | | | | | | |
| | nnDerW | 1--2 | 1--2 | 1--2 | 1 | 1 | |
| Latent period as unknown input | Identifiable | ρ, β, μ_1,μ_2 (All) | All | All | All | ρ | ψ, δ, γ_0, γ_i, γ_h |
| | Non Identifiable | | | | | α_e, α_i, ρ, β, μ | α, η |
| | Observable states | A, S, E | S, E, I | S, E, I | S, E, I | | |
| | Non Observable states | R | R | P | R, P | E, I, S, R | A, E, R, S |
| | D. M. | I | Q | Q, R, D | | P | D, H, I |
| | Observable unknown inputs | γ | v | γ, k | vi, vs, κ, vr_d1, vp_d1, vr_d2, vp_d2 | κ, κ_d1 | β, σ |
| | Non Observable unknown inputs | | | λ | vp, vr | | |
| | Known inputs | | | | | | |
| | nnDerW | 1--2 | 1--2 | 1--2 | 1--2 | 1 | 1 |
| Recovery rate as unknown input | Identifiable | ρ, β, γ (All) | All | All | All | | δ, ψ, σ |
| | Non Identifiable | | | | | α_e, α_i, μ, κ | α, η |
| | Observable states | A, S, E | S, E, I | S, E, I | S, E, I | | |
| | Non Observable states | R | R | P | R, P | E, I, S, R | A, E, R, S |
| | D. M. | I | Q | Q, R, D | | P | D, H, I |
| | Observable unknown inputs | μ_1, μ_2 | ψ | δ, k | vi, vs, ρ, β, vr_d1, vr_d2, vp_d1, vp_d2 | ρ_d1, β_d1 | β, γ_i, γ_0, γ_h |
| | Non Observable unknown inputs | | | λ | vp, vr | ρ, β | |
| | Known inputs | | | | | | |
| | nnDerW | 1--2 | 1--2 | 1--2 | 1--2 | 1 | 1 |
| Unknown inputs | Identifiable | | | | | | |
| | Non Identifiable | | | | | | |
| | Observable states | A, S, E | E, I, S | S, E, I | E, I, S | | |
| | Non Observable states | R | R | P | R, P | E, I, S, R | A, E, R, S |
| | D. M. | I | Q | Q, R, D | | P | I, H, D |
| | Observable unknown inputs | ρ, β, γ, μ_1, μ_2 | v, β, γ, ψ | k, α, β, δ, γ | α_e, α_i, ρ, β, μ, κ, vi, vs, vr_d1, vp_d1 | | β, δ, γ_0, γ_i, γ_h, ψ, σ |
| | Non Observable unknown inputs | | | λ | vr, vp | α_e, α_i, ρ, β, μ, κ | α, η |
| | Known inputs | | | | | | |
| | nnDerW | 1 | 1--2 | 1--4 | 1 | 1 | 1 |

## Table 2

| | | 39 | | 51 | |
|---|---|---|---|---|---|
| | | h=F, Di, De | h=Di, F | h=Di, F, De | h=Di, F |
| Constant parameters | Identifiable | μ_d, σ_d, γ_d | μ_d, σ_d, γ_d | β, γ, γ_d, μ_0, μ_d, μ_i, σ, σ_d | β, γ, γ_d, μ_0, μ_d, μ_i, σ, σ_d |
| | Non Identifiable | q, μ_i, σ, θ_e, θ_i, γ, β, β_d, ψ, ψ_e | q, μ_i, σ, θ_e, θ_i, γ, β, β_d, ψ, ψ_e | ψ, ψ_e, θ_e, θ_i, q, β_d | ψ, ψ_e, θ_e, θ_i, q, β_d, q |
| | Observable states | | De | S, E, I | S, E, I, De |
| | Non Observable states | E, I, R, S | E, I, R, S | R | R |
| | D. M. | De, Di, F | Di, F | Di, De, F | Di, F |
| | Observable unknown inputs | | | | |
| | Non Observable unknown inputs | | | | |
| | Known inputs | | | | |
| Transmission rate as unknown input | Identifiable | μ_d, σ_d, γ_d | μ_d, σ_d, γ_d | γ, γ_d, μ_0, μ_d, μ_i, σ, σ_d | γ, γ_d, μ_0, μ_d, μ_i, σ, σ_d |
| | Non Identifiable | q, μ_i, σ, θ_e, θ_i, γ, ψ, ψ_e | q, μ_i, σ, θ_e, θ_i, γ, ψ, ψ_e | ψ, ψ_e, θ_e, θ_i, q | ψ, ψ_e, θ_e, θ_i, q |
| | Observable states | | De | S, E, I | S, E, I, De |
| | Non Observable states | R | E, I, R, S | R | R |
| | D. M. | De, Di, F | Di, F | Di, De, F | Di, F |
| | Observable unknown inputs | | | β | β |
| | Non Observable unknown inputs | β, β_d | β, β_d | β_d | β_d |
| | Known inputs | | | | |
| | nnDerW | 1 | 1 | 1 | 1 |
| Latent period as unknown input | Identifiable | γ_d, μ_d, β, μ_i, γ | γ_d, μ_d, γ, μ_i, β | β, γ, γ_d, μ_0, μ_d, μ_i | β, γ, γ_d, μ_0, μ_d, μ_i |
| | Non Identifiable | q, θ_e, θ_i, β_d, ψ, ψ_e | q, θ_e, θ_i, β_d, ψ, ψ_e | ψ, ψ_e, θ_e, θ_i, q, β_d | ψ, ψ_e, θ_e, θ_i, q, β_d |
| | Observable states | E, I, S | S, E, I, De | S, E, I | S, E, I, De |
| | Non Observable states | R | R | R | R |
| | D. M. | De, Di, F | Di, F | Di, De, F | Di, F |
| | Observable unknown inputs | σ_d, σ | σ, σ_d | σ, σ_d | σ, σ_d |
| | Non Observable unknown inputs | | | | |
| | Known inputs | | | | |
| | nnDerW | 1--4 | 1 | 1--2 | 1--2 |
| Recovery rate as unknown input | Identifiable | μ_d, σ_d | μ_d, σ_d | β, μ_0, μ_d, μ_i, σ, σ_d | β, μ_0, μ_d, μ_i, σ, σ_d |
| | Non Identifiable | q, μ_i, σ, θ_e, θ_i, β, β_d, ψ, ψ_e | q, μ_i, σ, θ_e, θ_i, β, β_d, ψ, ψ_e, σ | ψ, ψ_e, θ_e, θ_i, q, β_d | ψ, ψ_e, θ_e, θ_i, q, β_d |
| | Observable states | | De | S, E, I | S, E, I, De |
| | Non Observable states | R | S, E, I, R | R | R |
| | D. M. | De, Di, F | F, Di | Di, De, F | Di, F |
| | Observable unknown inputs | γ_d, γ_d1, γ_d2 | γ_d, γ_d1, γ_d2 | γ, γ_d | γ, γ_d |
| | Non Observable unknown inputs | γ | γ | | |
| | Known inputs | | | | |
| | nnDerW | 1--2 | 1--2 | 1 | 1 |
| Unknown inputs | Identifiable | | | | |
| | Non Identifiable | | | | |
| | Observable states | S, E, I | De, S, E, I | S, E, I | S, E, I, De |
| | Non Observable states | R | R | R | R |
| | D. M. | De, Di, F | Di, F | Di, De, F | F, Di |
| | Observable unknown inputs | μ_d, σ_d, γ_d, μ_i, σ, γ, β | μ_d, σ_d, γ_d, μ_i, σ, γ, β | β, γ, γ_d, μ_0, μ_d, μ_i, σ, σ_d | β, γ, γ_d, μ_0, μ_d, μ_i, σ, σ_d |
| | Non Observable unknown inputs | q, θ_e, θ_i, β_d, ψ, ψ_e | q, θ_e, θ_i, β_d, ψ, ψ_e | ψ, ψ_e, θ_e, θ_i, q, β_d | ψ, ψ_e, θ_e, θ_i, q, β_d |
| | Known inputs | | | | |
| | nnDerW | 1 | 1 | 1 | 1 |

## Table 1

| | | | 18 | | | 31 | |
|---|---|---|---|---|---|---|---|
| | | | h= D | h=Q, D | h= D, I, Q | h=I | h=I, J |
| Constant parameters | | Identifiable | λ, σ, p | λ, σ, p | β, λ, σ, ρ, p, θ, εa, εi, γa | α, β1, σ, h, f | h, q, r, α, β1, β2, γ, φ, σ, f |
| | | Non Identifiable | β, εa, εi, γi, γa, γd, di, dd, θ, ρ | β, εa, εi, γi, γa, γd, di, dd, θ, ρ | γi, γd, di, dd | q, r, β2, γ, φ | |
| | | Observable states | | E, S | A, E, S | A, E, S | A, E, S |
| | | Non Observable states | A, I, R, Q, E, S | A, I, R | R | R, J | R |
| | | D. M. | D | Q, D | D, I, Q | I | I, J |
| | | Observable unknown inputs | | | | | |
| | | Non Observable unknown inputs | | | | | |
| | | Known inputs | | | | | |
| Transmission rate as unknown input | | Identifiable | λ, σ, p | λ, σ, p | λ, σ, ρ, p, θ, εa, εi, γa | β1, σ, h, f | β_1, β_2, ψ, σ, h, q, r, γ, f |
| | | Non Identifiable | εa, εi, γi, γa, γd, di, dd, θ, ρ | εa, εi, γi, γa, γd, di, dd, θ, ρ | γi, γd, di, dd | q, r, β2, γ, φ | |
| | | Observable states | | E, S, A, I | A, E, S | A, E, S, R, J | A, E, S |
| | | Non Observable states | A, I, R, Q, E, S | R | R | 0 | R |
| | | D. M. | D | Q, D | D, I, Q | I | I, J |
| | | Observable unknown inputs | | | β | α | α |
| | | Non Observable unknown inputs | β | β | | | |
| | | Known inputs | | | | | |
| | | nnDerW | 1 | 1 | 1 | 1 | 1 |
| Latent period as unknown input | | Identifiable | λ, p | p, λ | β, λ, ρ, p, θ, εa, εi, γa | α, σ, h, f | γ, σ, α, h, r, q, ψ, f |
| | | Non Identifiable | β, εa, εi, γi, γa, γd, di, dd, θ, ρ | β, εa, εi, γi, γa, γd, di, dd, θ, ρ | γi, γd, di, dd | q, r, γ, φ | |
| | | Observable states | | E, S, A, I | A, E, S | A, E, S | A, E, S |
| | | Non Observable states | A, I, R, Q, E, S | R | R | R, J | R |
| | | D. M. | D | Q, D | D, I, Q | I | I, J |
| | | Observable unknown inputs | σ | σ | σ | β1 | β1, β2 |
| | | Non Observable unknown inputs | | | | β2 | |
| | | Known inputs | | | | | |
| | | nnDerW | 1 | 1 | 1 | 1 | 1 |
| Recovery rate as unknown input | | Identifiable | λ, σ, p | λ, σ, p | β, λ, σ, ρ, p, θ, εa, εi | α, β1, σ, h, f | β_1, β_2, σ, α, h, r, q, ψ, f |
| | | Non Identifiable | β, εa, εi, di, dd, θ, ρ | β, εa, εi, di, dd, θ, ρ | di, dd | h, q, r, β2, φ | |
| | | Observable states | | E, S | A, E, S | A, E, S | A, E, S |
| | | Non Observable states | A, I, R, Q, E, S | A, I, R | R | J, R | R |
| | | D. M. | D | Q, D | D, I, Q | I | I, J |
| | | Observable unknown inputs | γi_d1, γa_d1, γd_d1 | γi_d1, γa_d1, γd_d1 | γa, γi_d1, γi_d2, γd_d1, γd_d2 | | γ |
| | | Non Observable unknown inputs | γi, γa, γd | γi, γa, γd | γi, γd | γ | |
| | | Known inputs | | | | | |
| | | nnDerW | 1 | 1 | 1--2 | 1 | 0--1 |
| Unknown inputs | | Identifiable | | | | | |
| | | Non Identifiable | | | | | |
| | | Observable states | A, I | E, S, A, I | A, E, S | A, E, S, R, J | A, E, S |
| | | Non Observable states | R, Q, E, S | R | R | | R |
| | | D. M. | D | Q, D | D, I, Q | I | I, J |
| | | Observable unknown inputs | λ, σ, p, β, εa, εi, γa, θ, ρ | λ, σ, p, β, εa, εi, γa, θ, ρ | β, λ, σ, ρ, p, θ, εa, εi, γa | α, β1, σ, f, q, r, β2, γ, φ | h, q, r, α, β1, β2, γ, φ, σ, f |
| | | Non Observable unknown inputs | γi, γd, di, dd | γi, γd, di, dd | γi, γd, di, dd | h | |
| | | Known inputs | | | | | |
| | | nnDerW | 1 | 1 | 1 | 1 | 1 |

## Table 2

| | | | 17 | | | 32 | 33 |
|---|---|---|---|---|---|---|---|
| | | | h=D | h=Q, D | h= D, I, Q | h=WL | h=L, Q |
| Constant parameters | | Identifiable | q, ε | d, q, qt, β, γ | d, q, qt, β, ε, γ | W, β, α, μ, γ | γ, β1, η, δ, ξ, θ2, ε, θ1, α1, α2 |
| | | Non Identifiable | β, γ, d, qt | | | | |
| | | Observable states | | E, I, S | E, S | I, S, L | E, I, S |
| | | Non Observable states | E, I, R, Q, S | R | R | R | R |
| | | D. M. | D | Q, D | Q, D, I | | L, Q |
| | | Observable unknown inputs | | | | | |
| | | Non Observable unknown inputs | | | | | |
| | | Known inputs | | | | | |
| Transmission rate as unknown input | | Identifiable | q, ε | All | All | All | All |
| | | Non Identifiable | γ, d, qt | | | | |
| | | Observable states | | E, I, S | E, S | I, S, L | E, I, S |
| | | Non Observable states | E, I, R, Q, S | R | R | R | R |
| | | D. M. | D | Q, D | Q, D, I | | L, Q |
| | | Observable unknown inputs | | β | β | β | β |
| | | Non Observable unknown inputs | β | | | | |
| | | Known inputs | | | | | |
| | | nnDerW | 1 | 1 | 1 | 1 | 1 |
| Latent period as unknown input | | Identifiable | q | All | All | All | All |
| | | Non Identifiable | β, γ, d, qt | | | | |
| | | Observable states | | E, I, S | E, S | I, S, L | E, I, S |
| | | Non Observable states | E, I, R, Q, S | R | R | R | R |
| | | D. M. | D | Q, D | Q, D, I | | L, Q |
| | | Observable unknown inputs | ε | ε | ε | W | ε |
| | | Non Observable unknown inputs | | | | | |
| | | Known inputs | | | | | |
| | | nnDerW | 1--2 | 1--2 | 1--2 | 1 | 1 |
| Recovery rate as unknown input | | Identifiable | q, ε | All | All | All | All |
| | | Non Identifiable | β, d | | | | |
| | | Observable states | | E, I, S | E, S | I, S, L | E, I, S |
| | | Non Observable states | E, I, R, Q, S | R | R | R | R |
| | | D. M. | D | Q, D | Q, D, I | | L, Q |
| | | Observable unknown inputs | γ_d1, qt_d1, γ_d2, qt_d2 | γ, qt | γ, qt | γ | α1, α2 |
| | | Non Observable unknown inputs | γ, qt | | | | |
| | | Known inputs | | | | | |
| | | nnDerW | 1--2 | 1--2 | 1--2 | 1 | 1 |
| Unknown inputs | | Identifiable | | | | | |
| | | Non Identifiable | | | | | |
| | | Observable states | | E, I, S | E, S | | E, I, S |
| | | Non Observable states | E, I, R, Q, S | R | R | I, S, L | R |
| | | D. M. | D | Q, D | Q, D, I | R | L, Q |
| | | Observable unknown inputs | q, ε | d, q, qt, β, ε, γ | d, q, qt, β, ε, γ | W, β, α, μ, γ | γ, β1, η, δ, ξ, θ2, ε, θ1, α1, α2 |
| | | Non Observable unknown inputs | β, γ, d, qt | | | | |
| | | Known inputs | | | | | |
| | | nnDerW | 1 | 1 | 1 | 1 | 1 |






\bibliographystyle{elsarticle-num-names}
\bibliography{biblio_ident.bib,biblio_table.bib,biblio_models.bib,biblio_Intro.bib}







\end{document}


\maketitle

The results of the observability and structural identifiability study carried out through STRIKE-GOLDD and ObservabilityTest, have been included in table form. Each group of rows represents a special case studied: constant parameters (i.e. as found in its original source), considering the transmission rate, the latent period or the recovery rate as time-dependent inputs and, finally, all parameters as unknown functions. 
These large groups of rows are in turn divided to give more detailed information: identifiable and non-identifiable parameters, observable and non-observable states, directly measured states, unknown observable and unobservable input functions, known inputs and, finally, the number of non-negative derivatives considered (nnDerW).

Other observations about the table: the suffix $\_$\textit{d number} means the nth derivative of an unknown function (ej. $\beta\_d1$); the missing blank in the SEIR 18 and SEIR 8 means that this case has already been studied with the original considerations of the model and  SIR 29 and 30 models have only been studied in their original form because the common parameters of the SIR models are within $R_0$.